\documentclass[pre,eqsecnum,preprint,showpacs,floatfix]{revtex4}
\usepackage{amssymb,amsmath,bm,graphicx}
\font\bba=msbm10 scaled 1080
\font\bbb=msbm8 
\font\bbc=msbm6 
\newfam\bbfam
\textfont\bbfam=\bba
\scriptfont\bbfam=\bbb
\scriptscriptfont\bbfam=\bbc

\begin{document}
\title{An accurate equation of state for the one component plasma in the 
low coupling regime.}
\author{Jean-Michel Caillol}
\affiliation{Laboratoire de Physique Th\'eorique \\
CNRS (UMR 8627), B\^at. 210\\
Universit\'e de Paris-Sud \\
91405 Orsay Cedex, France}
\email{Jean-Michel.Caillol@th.u-psud.fr}
\author{Dominique Gilles}
\affiliation{CEA/DSM/Institut de Recherche sur les lois Fondamentales de l'Univers\\
CE Saclay\\
91191 Gif sur Yvette Cedex, France}
\email{Dominique.Gilles@cea.fr}
\date{\today}
\begin{abstract}
An accurate equation of state of the one component plasma is obtained in the low coupling 
regime $0 \leq \Gamma  \leq 1$. The accuracy results from a smooth combination of the well-known hypernetted chain integral
equation, Monte Carlo simulations and asymptotic analytical expressions of the excess internal energy $u$. In particular, 
special attention has
been brought to describe and take advantage of finite size effects on Monte Carlo results to get the thermodynamic limit of $u$.
This combined approach reproduces very accurately
the different plasma correlation regimes encountered in this range of values of $\Gamma$.
This paper extends to low $\Gamma$'s an earlier Monte Carlo simulation study devoted to strongly coupled systems for
$1 \leq \Gamma  \leq 190$ (\mbox{J.-M. Caillol}, \mbox{J. Chem. Phys. } \textbf{111}, 6538  (1999)). 
Analytical fits of  $u(\Gamma)$ in the range $0 \leq \Gamma  \leq 1$ are provided with a precision that we claim to be 
not smaller than $p= 10^{-5}$. HNC equation and exact asymptotic expressions  are shown to 
give reliable results for $u(\Gamma)$ only in narrow $\Gamma$ intervals, i.e. $0 \leq \Gamma  \lesssim 0.5$ and
$0 \leq \Gamma  \lesssim 0.3$ respectively. 
\end{abstract}
\pacs{52.65.-y, 52.25.Kn, 52.27.Aj}
\maketitle
\newpage
\section{Introduction}
\label{intro}
The aim of this paper is to obtain the equation of state (EOS) of a plasma in the low coupling regime with a high precision.
In this regime standard Monte Carlo (MC) and Molecular Dynamics
simulations techniques must be handled with care due to huge finite size effects  and, in the other hand,
the ideal gas approximation or more elaborated analytical expressions commonly used are valid only but asymptotically,
for very small values of the coupling parameters. 
Such thermodynamic conditions  are relevant for many astrophysical or laboratory plasmas hydrodynamics applications.

However we shall restrict ourselves to the well known one-component plasma (OCP) model, which consists of identical
point ions with number density  $n$, charges $Ze$, moving 
in a neutralizing background, electrons for instance, where $n=N/\Omega $, $N$ number of particles,
$\Omega$ volume of the system\cite{Hansen1}.
In the very low coupling regime, the virial expansion supplemented  by well documented resummation methods, 
as the well-known Debye-H\"uckel (DH) theory \cite{Debye} and its extensions (see e.g. Cohen \cite{Cohen} and,
more recently, Ortner  \cite{Ortner} expansions for instance) give reliable results.  In the low to intermediate coupling
regimes the HyperNetted Chain (HNC) integral equation \cite{Ng} must be solved numerically. Finally,  in the strong correlation 
regime, the OCP has also been extensively studied 
by Monte Carlo and Molecular Dynamics simulations for three decades, see e.g.
\cite{Teller,Hansen1,Pollock,DeWitt,DeWitt2, Caillol1, Caillol2} and references cited herein.

In the more recent of these references one of us has determined the thermodynamic limit of the excess internal 
energy per particle $u_{N=\infty}$ of the OCP
with a high precision by means of MC simulations in the canonical ensemble within hyperspherical boundary 
conditions \cite{Caillol1, Caillol2} for 
$1 \leq \Gamma  \leq 190 $. We recall that in the thermodynamic limit, i.e. for an infinite system of particles,
the thermodynamics properties of the model depend solely on the
coupling parameter $\Gamma=\beta (Z e)^{2}/ a_{i}$ ($\beta=1/kT$, $k$ Boltzmann constant, $T$ temperature,
and  $a_{i}$ the ionic radius defined by $4\pi n a_{i}^{3}/3=1$), whereas, for a finite sample, an additional dependance
on the number of particles $N$ remains.
In paper\ \cite{Caillol2}, henceforth to be referred to as "I",  special attention has
been brought to describe and take advantage of such finite size effects on the energy $u_{N}(\Gamma)$
to get its thermodynamic limit $u_{N=\infty}$, using all facilities of work stations available at that time.

Recently we have also performed extensive MC simulations of the related Yukawa One-Component Plasma (YOCP), i.e. a
system made  of $N$ identical point charges $Z e$ interacting via an effective Yukawa pair-potential 
$v_{\alpha}(r)=(Z e)^{2}\exp(-\alpha r)/r$,
where $\alpha$ is the so-called screening parameter \cite{CGYOCP}, not to be discussed however in this work.
For the OCP and the YOCP as well, in the low $ \Gamma$ regime, the Debye length 
(i.e. the correlation length associated with charge fluctuations) becomes of the order or much larger than the size of the
simulation box, yielding huge finite size effects on  $u_{N}(\Gamma)$.
Therefore, despite these numerous studies and amount of work it appears that hydrocode applications using
the combination of data
bases and fits coming from various techniques can be affected by numerical instabilities in the transition regime, around
$\Gamma =1$. With nowadays computers it is now possible to explore this range of small $\Gamma$ values
with the help of performant simulation techniques and to obtain such precise results so that they can be considered as
the \textit{reference} ones to be used in many applications dealing with degenerate astrophysical or laboratory plasmas.
We also examine carefully in this paper
the connection between MC and first principle analytical or HNC results for $ \Gamma  \leq 1$. We have thus
explored and precised the domain of validity of 
each of these methods. It turns out to be necessary to combine all of these approaches to obtain 
a continuous representation of  $u_{N=\infty}(\Gamma)$ in the range $0 \leq \Gamma  \leq 1$. Finally
we extract from  these combined approaches the best possible  analytical representation for 
$u_{\infty}(\Gamma)$.

Our paper is organized as follows.
Next section is devoted to a brief presentation of the main features of  low $\Gamma$ expansions 
(Section \ref{Ortner}), the HNC integral equation (Section \ref{HNC}) and the rather unusual but efficient MC 
technique used in this paper (Section \ref{MC}).
Note that we have redone, by passing, extremely accurate HNC calculations and obtained new fits of HNC data, presented in
Section \ref{HNC}. In Section \ref{MCdat} we present and discuss  our MC  simulations.
Fits of the data are described in details and widely illustrated.
Finally conclusions are drawn in Section \ref{Conclusion}.

\section{Low $\Gamma$ calculation Methods}
\label{par1}
\begin{figure}
\includegraphics[angle=0,scale=0.4]{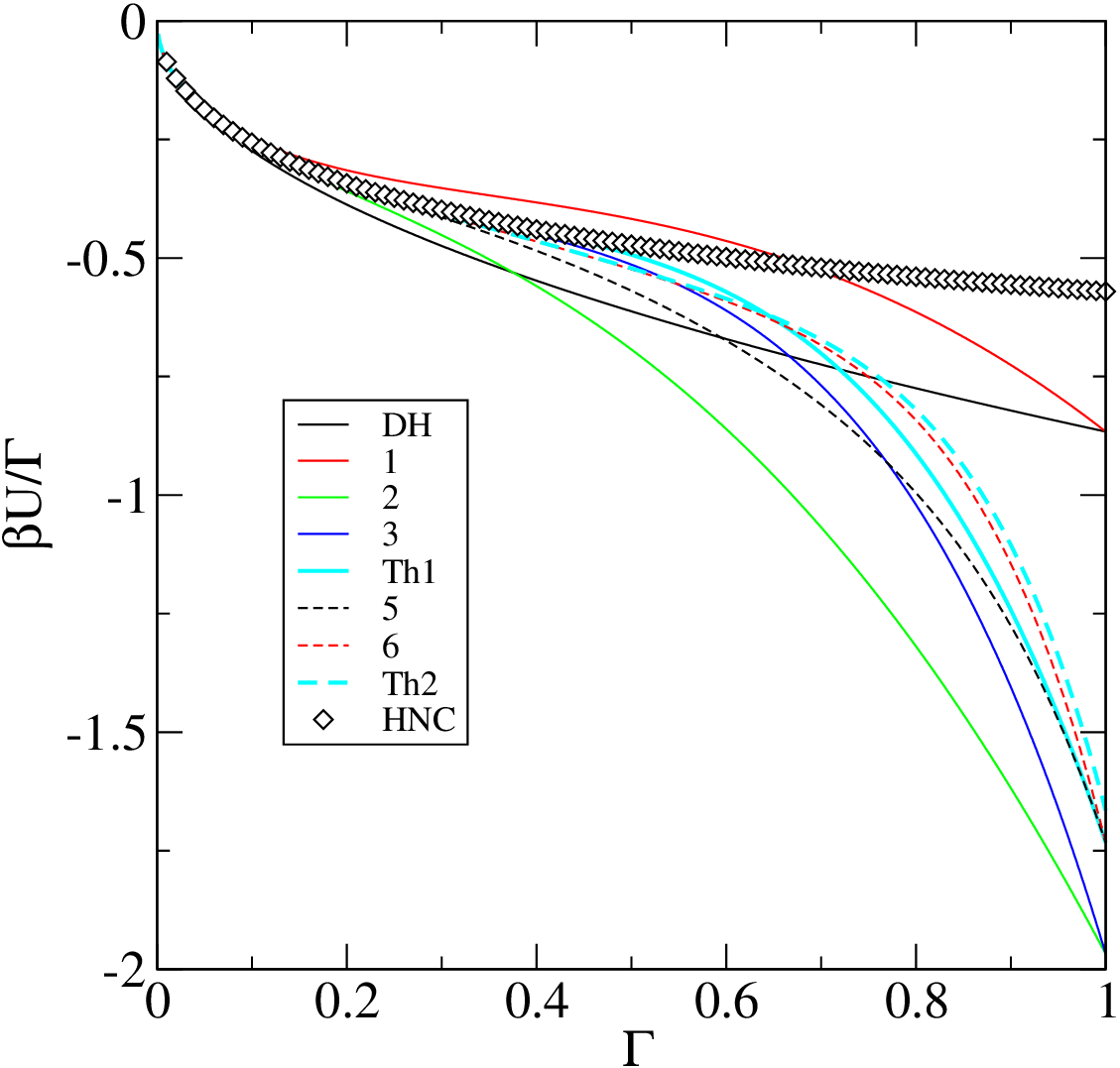}
\caption{\label{fig_U_Ortnert} 
Reduced excess energy $\beta u /\Gamma$ versus $0\leq \Gamma \leq 1$. Diamonds: HNC,
black solid line: DH, thick cyan solid line: Th1 approximation\ (\ref{a}), 
thick cyan dashed line: Th2 approximation\ (\ref{b}), other curves
represent the successive orders of expansion\ (\ref{eqnortner}).
}
\end{figure}
\begin{figure}
\includegraphics[angle=0,scale=0.4]{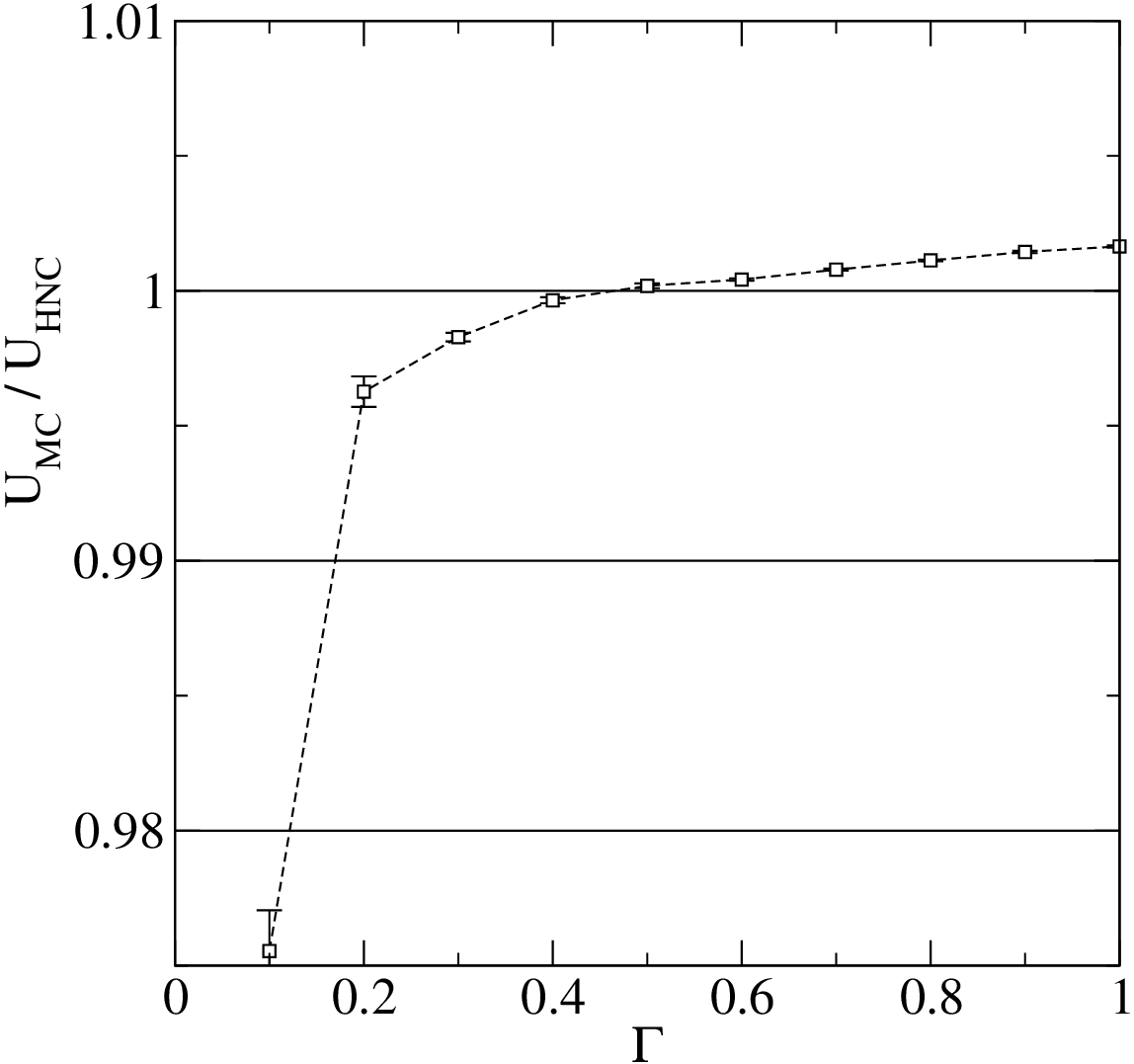}
\caption{\label{fig_RAP} 
Ratio of MC excess energies to HNC results versus $\Gamma$ in the low coupling regime.
}
\end{figure}

The interval $0 \leq \Gamma  \leq 1$ covers various correlation regimes from no correlation
($\Gamma =0 $, i.e. the ideal gas) to
an intermediate correlated regime ($\Gamma =1 $, no oscillation or structure in the pair correlation functions).
In any case,  the long-range nature of the interaction potential between 
two ionic charges causes  Mayer graphs to diverge \cite{Hansen1}.
A field theoretical diagrammatic representation of  cluster integrals  has been proposed recently in \cite{Ortner} to 
avoid complicated chain resummations in an attempt to treat the $\Gamma$ expansion of the classical Coulomb system
in a more controlled and systematic way.
In this interesting paper the final expansion obtained by the author improves earlier and seminal 
analytical results of Cohen \textit{et al.} \cite{Cohen, Debye} obtained by traditional diagrammatic 
expansions and resummations. From these theoretical analysis it turns out that the physics in this small
interval $0 \leq \Gamma  \leq 1$ is extremely complicated and exhibits many different correlation regimes, even more
than in the widely studied region  $1 \leq \Gamma  \leq 190$ \cite{Teller,Hansen1,Pollock,DeWitt,DeWitt2, Caillol1, Caillol2}.
The low $\Gamma$ expansions obtained by Cohen \textit{et al.} and Ortner for $u_{\infty}(\Gamma)$ converge
to the HNC results 
only for  $0 \leq \Gamma \leq 0.2 $ as apparent in figure~\ref{fig_U_Ortnert}. For higher values of $\Gamma$ these asymptotic
expressions do not seem to converge at all and, moreover, the high order terms of the expansions do not improve
the results of the lower orders.  
Anticipating the results of sections\ \ref{HNC} and\ \ref{MC} and, as can be observed in
figure~\ref{fig_RAP},  the HNC data 
deviate from our MC results as soon as  $\Gamma  \geq 0.5$. It results from this sketchy discussion that we must 
distinguish three different regimes of correlations in the interval $0 \leq \Gamma  \leq 1$, and we confess that this complexity
motivated the present study.

\subsection{Cohen and Ortner analytical expansions}
\label{Ortner}
In ref.~ \cite{Ortner} Ortner has developed an effective method based on the 
Hubbard-Stratonovich (HS) transformation
and field theoretical  approaches  to calculate the free energy 
of classical Coulomb systems in the low  $\Gamma$ regime  \cite{Hubbard,Strato,Caillol3}.
The HS transform  was used to obtain the EOS  of a classical plasma and notably that of the OCP.
The non-trivial  part of the Helmholtz free-energy density was derived up to order 
$\Gamma^6$, improving on the previous results of Cohen \textit{et al.} at  order $\Gamma^{\frac{9}{2}}$, obtained 
by a method of resummation of diverging diagrams. The author gives an analytical
representation
of the excess internal energy $\beta u$ of the OCP, valid at low  $\Gamma$, without however any estimation of the error.
It reads as,
\begin{subequations}
\label{eqnortner}
\begin{align}
\beta u (\Gamma)&= p_0 \Gamma^{3/2} + p_1 \Gamma^{3} \ln \Gamma + p_2 \Gamma^{3}  
             +                   p_3 \Gamma^{9/2} \ln \Gamma + p_4 \Gamma^{9/2}  \label{a} \\
              & +    p_5 \Gamma^{6} \ln^{2}\Gamma                       
            +p_6 \Gamma ^{6}  \ln \Gamma + p_7 \Gamma ^{6}       \label{b}   \, 
\end{align}
\end{subequations}
with the constants, 
$p_0 = - \sqrt{3}/2$, $p_1=-9/8$, $p_2=-(9 \ln3)/8- 3\,C_E/2+1$, $p_3=-(27\sqrt{3}) /16$, $p_4=0.2350$, 
$p_5=-81/16$, $p_6=-2.0959$, $p_7=0.0676$ and $C_E=0.57721566$ the Euler  constant.
Expression\ \ref{eqnortner} (to be referred to as Th2 henceforth)
 improves on that given by Cohen \textit{et al.}(to be referred to as Th1 henceforth) \cite{Cohen},
 which corresponds to
line \ref{a}, while  the additional  terms are those of  line \ref{b}.
We recognize that the first term ($- \sqrt{3}\Gamma^{3/2}/2$) is exactly the well known Debye-H\"uckel (DH) 
contribution.
Figure \ref{fig_U_Ortnert}  displays the results of the reduced excess energy $\beta u /\Gamma$ versus $\Gamma$ 
at successive orders in the $\Gamma$-expansion~\ref{eqnortner}. A close examination of the figures reveals that
the DH approximation is nearly exact up to $\Gamma=0.05$, in the sense that higher order contributions do not change 
the result. A comparison with HNC results,  which are supposed to be nearly exact 
at least up to $\Gamma=0.5$ (this point will be  fully discussed in next section), shows the convergence of the expansions 
Th1 and Th2 to  HNC at $\Gamma \leq 0.3$ and  $\Gamma \leq 0.2$ respectively.
However we do not observe any trend of convergence of these expansions for $\Gamma \geq 0.4$. 
We also notice that the additional terms given by Ortner (cf equation\ \ref{b}) lead to an oscillatory behavior rather than 
to an improved convergence radius. We suspect some misprints in the reported  $p_n$ for $n=5, 6, 7$ since the $\Gamma$
functional  $\Gamma$ dependence of\ \ref{eqnortner} is  undoubtedly correct.
\subsection{HNC method and fits}
\label{HNC}
\begin{figure}
\includegraphics[angle=0,scale=0.4]{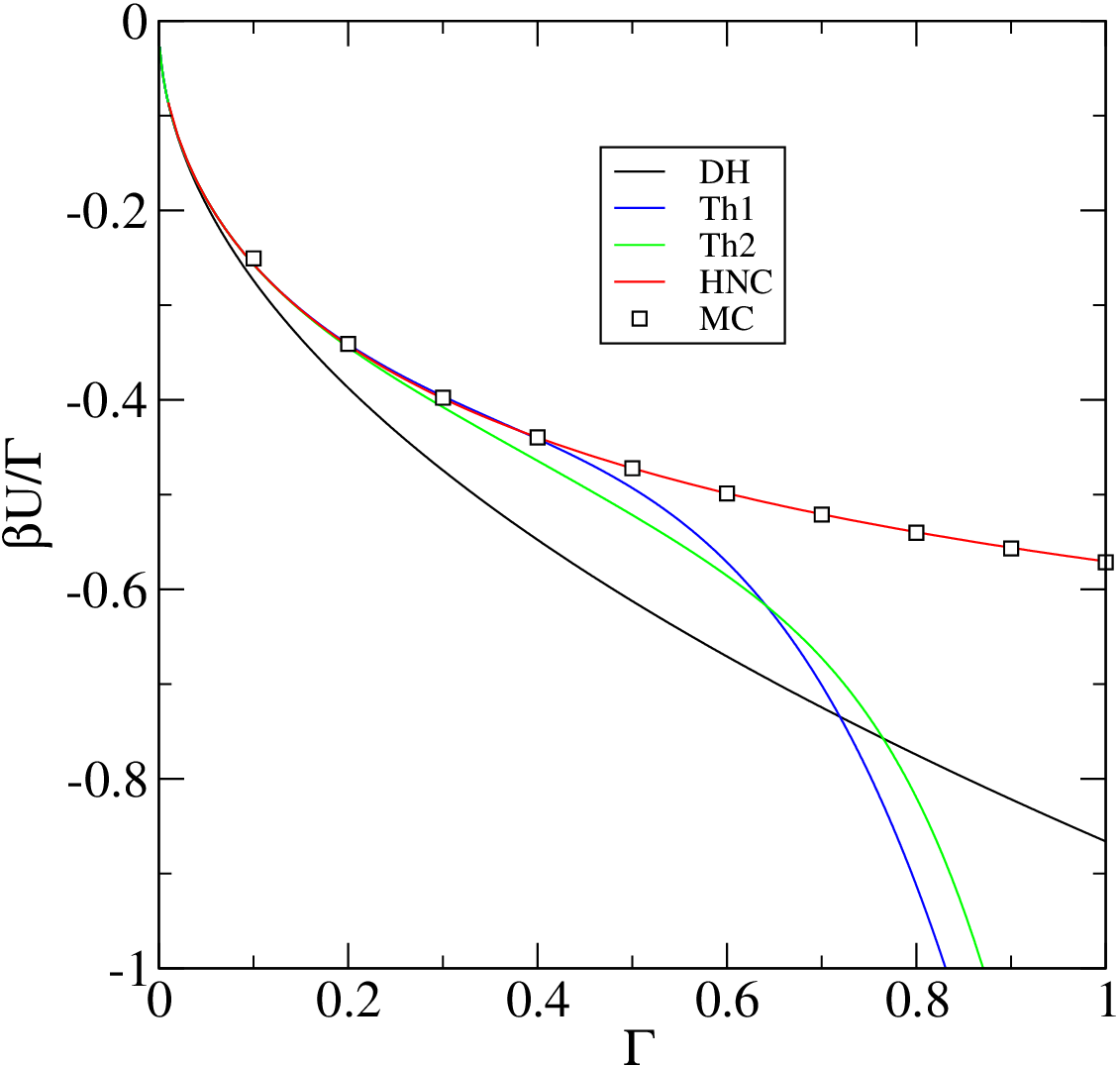}
\caption{\label{fig_COMPARAISON_U} 
Reduced excess energy $\beta u /\Gamma$ versus $ \Gamma $. 
Squares: MC (the symbols are larger than error bars),
black line: DH,
red line: HNC,
blue line: Th1 approximation\ (\ref{a}), 
green line: Th2 approximation\ (\ref{b}).
}
\end{figure}
\subsubsection{Method}
\label{HNCmeth}

\begin{figure}
\includegraphics[angle=0,scale=0.4]{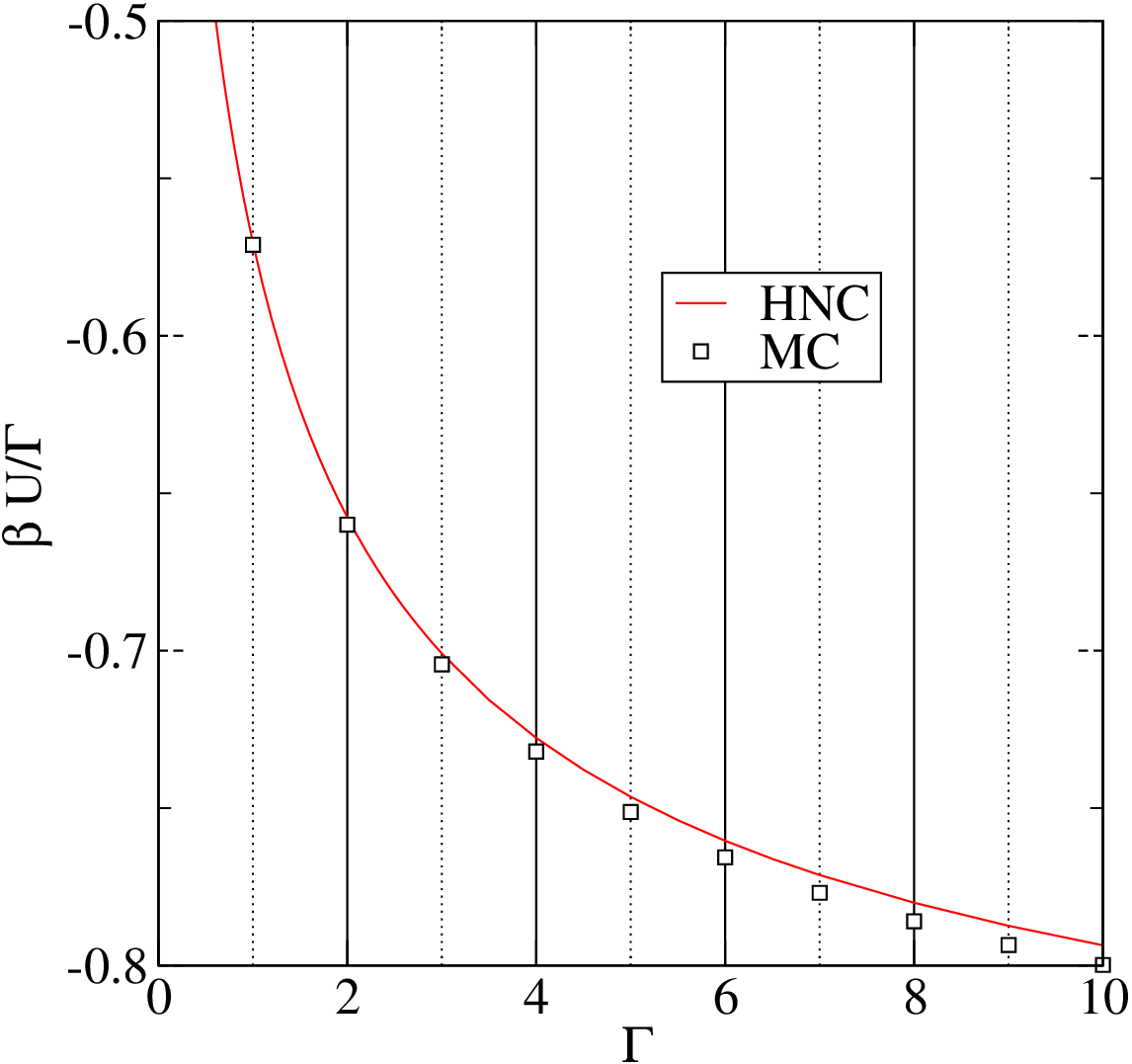}
\caption{\label{fig_HORREUR_DE_HNC} 
Comparison between HNC (red line) and MC data (squares, present work and previous
results, see ref.\cite{Caillol2}) for the reduced excess energy  $\beta U /\Gamma$
 versus $1\leq \Gamma \leq 10$.}
\end{figure}
We have redone high precision HNC calculations for a hundred of values of $\Gamma$ in the range $(0,1)$ (see 
figures~\ref{fig_U_Ortnert} and~\ref{fig_COMPARAISON_U} ); additional 
calculations were also done for some higher values of the coupling parameter, in the range $1 \leq \Gamma \leq 10$,
see figure~\ref{fig_HORREUR_DE_HNC}.
We  used  the Ng method\cite{Ng} with the following control parameters:
the pair correlation  functions (direct and non-direct respectively) $c(r)$ and $h(r)$,
as well as their Fourier transforms $\widetilde{c}(k)$
and  $\widetilde{h}(k)$, were tabulated on  grids of $N=2^{M}$ points with $M=20$ in order to make use of
fast Fourier transforms with intervals  of $\Delta r=0.001$ and  $\Delta k=2 \pi/N \simeq 6 10^{-3}$
in direct and Fourier space respectively. The dimensionless energies were computed
according the formulae\ \cite{Hansen1}
\begin{subequations}
\begin{eqnarray}
\label{Ur}
\dfrac{\beta u^{(r)}}{\Gamma} &=&\dfrac{3}{2} \; \int_{0}^{\infty}\; dr\; r h(r)  \; , \\
\label{Uk}
\dfrac{\beta u^{(k)}}{\Gamma} &=&\dfrac{3}{2 \; (2 \pi)^{2}} \; \int_{0}^{\infty}\; dk \; \widetilde{h}(k)  \; ,
\end{eqnarray}
\end{subequations}
where the distances "$r$" are measured in the units of the ionic radius $a_{i}$ and the wave
numbers $k$ in units of $a_{i}^{-1}$. The comparison of these two estimations $u^{(r)}$ and $u^{(k)}$ of the energy,
which of course should be equal, give an idea on the relative
precision of the numerical resolution of HNC, typically  about $10^{-12}$ at $\Gamma=0.01$
and $10^{-13}$ at $\Gamma \geq 0.1$. Another usefull test is to check the Stillinger-Lovett (SL)  sum rules
;recall briefly the two first SL rules (the third one should not be  satisfied by HNC\cite{Hansen1})
 \begin{subequations}
\begin{eqnarray}
\label{L1}
3 \; \int_{0}^{\infty}\; dr\; r^{2} h(r) &=& -1   \; , \\
\label{L2}
3 \; \int_{0}^{\infty}\; dr\; r^{4} h(r) &=& \dfrac{2}{\Gamma}   \; . 
\end{eqnarray}
 \end{subequations}
With the control parameters given above the SL rules were satisfied with a relative precision of  about $10^{-13}$.
\subsubsection{Fits}
\label{fitHNC}
\begin{table}
\caption{\label{coef4} First five  Cohen-Ortner coefficients (cf Eq.~(\ref{eqnortner}), first line)
compared to the correspondent coefficients of the fits of the 
energy $\beta u(\Gamma)/\Gamma$ for HNC and MC data.
Second line :  HNC, 7 parameters, $p0=-\sqrt{3}/2$ fixed to its DH value.
Third line     :  HNC, 8 parameters.
Last line      :  MC data in the range $0.4 \leq \Gamma \leq 1$ , 5 parameters ($p_5=p_6=p_7=0$).
}
\begin{tabular}{|l|l|l|l|l|l|l|l|l|} 
\hline
$p0$ &  $p1$  &   $p2$    &  $p3$   & $p4$  & $Method$ \\ \hline
$-0.8660254038$&$-1.1250000000$&$-1.1017662315$&$-2.9228357378$&$0.2350000000$& $Ortner$\\ \hline
$-0.8660254038$&$-1.1127645260$&$-1.0636075255$&$-3.1960177420$&$-1.4236810385$ & $HNC-DH$\\ \hline
$-0.8658509448$&$-1.0967358264$&$-1.0224523661$&$-2.9765709164$&$-1.1861133643$& $HNC$   \\ \hline
$-0.8409025523$&$-0.5198391670$&$-0.0001985314$&$-0.1402132305$&$0.2697081277$&$MC    $\\ \hline
\end{tabular}
\end{table}
\begin{table}
\caption{\label{coef567} Same as in Table\ \ref{coef4}} for the last 3 parameters $p_5$, $p_6$, $p_7$
of the fit of HNC data.
\begin{tabular}{|l|l|l|l|l|l|l|l|l|} 
\hline
$p5$    &  $p6$   & $p7$ & $Method$ \\ \hline
$-5.062500000$&$-2.0959000000$&$0.0676000000$ & $Ortner$\\ \hline
$ 0.5868725967$&$-2.1982700902$&$2.7828599024$ & $HNC-DH$\\ \hline
$ 0.5093239388$&$-1.9531860886$&$2.5039620685$ & $HNC$   \\ \hline
\end{tabular}
\end{table}
\noindent
We used the functional form of Ortner asymptotic expression\ \ref{eqnortner} to fit the HNC data for
$ \beta u/\Gamma$ in the interval $0 \leq \Gamma  \leq 1$. We are left with a eight parameters
fit (i.e. the $p_i$  for $i=0, \ldots 7$) or a seven parameters fit, if $p_0$ is fixed to its Debye value 
$p_0=-\sqrt{3}/2$. The values found for the $p_i$ are given in the Tables\ \ref{coef4} and\ \ref{coef567}.
 For the eight parameters
fit the maximum deviation of the fit from the HNC data is $7.3\,10^{-7}$ with a mean deviation of 
$1.9\, 10^{-7}$, while for the seven parameters fit these deviations are  $1.3  \,10^{-6}$
and $3.3 \, 10^{-7}$ respectively.
Some comments are in order.
\begin{itemize}
\item Firstly, for $\Gamma \leq 0.1$ the estimations of $ \beta u/\Gamma$ in the framework of
         HNC, Cohen \textit{et al.} and Ortner
        theories all coincide with an absolute precision of the order  of $1.  10^{-4}$, as apparent in table\ \ref{compar}. 
        These conclusions are also true for DH approximation.
\item The agreement between HNC energies and that predicted by Cohen \textit{et al.} expression  (cf ``Th1'' in figure\ 
         \ref{fig_COMPARAISON_U} and table\ \ref{compar}) differ by less than  $2.  10^{-3}$ in the range 
         $0 \leq\Gamma \leq 0.3$. Note that the apparent discrepancies between the $p_i$ of the fit of HNC and the ''exact''
         coefficients of Cohen expansion do not spoil the excellent agreement between the two approaches.
\item The agreement between HNC energies and that predicted by Ortner \textit{et al.} expression  (cf ``Th2'' in figure\ 
         \ref{fig_COMPARAISON_U} and table\ \ref{compar}) differ by less than  $2.  10^{-3}$ in the range 
         $0 \leq\Gamma \leq 0.2$.
\end{itemize}
From these remarks we conclude that HNC is, as expected, exact in the low coupling regime at least up to $\Gamma=0.3$.
Moreover DH theory cannot be trusted for  $\Gamma \geq 0.1$, Cohen \textit{et al.} expression can be used confidently
as it stands  for  $\Gamma \leq 0.3$  and, unexpectedly, the additional orders in the asymptotic expression  
obtained by Ortner  do not improve,  unfortunately, on Cohen results.
 We suggest to reexamine the details of the calculations of reference\ 
\cite{Ortner}. The functional Dependance in $\Gamma$ of equation\ (\ref{eqnortner}) is probably correct but misprints in one of
the $p_i$ for either $i=5,6$ or $i=7$ are likely.
\subsection{MC theoretical background}
\label{MC}
MC simulations are not well adapted  to the low coupling regime for two reasons. First, since the configurational
energies  are small,  the convergence of the MC process is slow. Secondly, in the case of the OCP
considered here, the Debye length $\lambda_D=1/\sqrt{3 \Gamma}$ diverges as $\Gamma \to 0$ and thus becomes larger
than the (finite) size of the simulation box, with entails severe finite size effects.
To use the MC  method for obtaining very precise results for the OCP in the range of $0 \leq \Gamma  \leq 1$ is therefore
a real challenge. Some comments on our methodology seem to us worthwhile.

Our simulations were performed in the canonical ensemble within hyperspherical boundary conditions. 
The particles are thus confined on the surface of a $4D$ sphere ${\cal S}^{3}$ of radius $R$ and the plasma pair potential
between ions is simply the Coulombic interaction in this geometry.
The latter has a simple analytical expression  which allows high precision
computations in contrast with the usual technique of Ewald summations where the potential is poorly 
determined at short distances. The theoretical background  of this method has been already
described in details in previous works\ \cite{Caillol1, Caillol2} and will not be rediscussed here.
We only extract from
these previous theoretical considerations  the following point. It turns out that DH equation
(i.e. Helmoltz equation) can be solved analytically in 
${\mathcal S}^{3}$ which yields the exact finite size dependence of the excess internal energy in this 
approximation and therefore in the low coupling limit. One finds that at the leading order
\begin{equation}
\label{scaling}
u_{N}\left( \Gamma \right) - u_{\infty}\left( \Gamma \right) \sim N^{-2/3} \;  \mathrm{ for } \; \Gamma \to 0  \; \mathrm{ and }
\; N \to \infty \; .
\end{equation}
Of course this behavior in only asymptotic and sub-leading terms in 
$ \left[ N^{-2/3}\right] ^{2}$,  $\left[ N^{-2/3}\right] ^{3}$ must be taken into account if $N$ is not large enough.
 For couplings $\Gamma \geq 3$ we shown in paper I that 
we  rather have   $u_{N}\left( \Gamma \right) - u_{\infty}\left( \Gamma \right) \sim N^{-1}$.
This remark yields the correct procedure : for a given parameter $\Gamma$ perform MC simulations for
different number of particles $N$ and take advantage of the scaling relation\ \ref{scaling} to obtain the thermodynamic limit 
$ u_{\infty}\left( \Gamma \right)$. The estimation of the  statistical errors on the $ u_{N}\left( \Gamma \right)$ and
the extrapolated thermodynamic limit  $u_{\infty}\left( \Gamma \right)$  is 
also described in details in I. 
However, by contrast with refs\ \cite{Caillol1, Caillol2} devoted to the strong correlation regime ($1 \leq \Gamma \leq190 $), 
present work only the small couplings are considered.
In order to test the validity of HNC, notably in the range
$0.3 \leq \Gamma \leq 1$ with an error of $\sim 1. 10^{-4}$ we were led to perform huge Markov chains and 
consider very large systems up to $N=51200$  particles in order to reach the scaling region where\ \ref{scaling}
applies. Since HNC and Cohen asymptotic forms for $u$ differ by less than  $\sim 1.  \,10^{-4}$ in the range 
$0 \leq \Gamma \leq 0.3$ we can claim (as will be discussed in details below) an overall maximum error of $\sim 1. \, 10^{-4}$
for the dimensionless $\beta u/\Gamma$ in the whole interval $0 \leq \Gamma \leq 1$.

Some additional simulations 
in the transition region to high correlation regime $1 \leq \Gamma \leq 10$ were also performed to make 
contact with our previous results.
\begin{table*}
\caption{\label{compar} Minus the dimensionless energy $\beta u_{N}(\Gamma)/\Gamma$ of
the OCP as a function of $\Gamma$ for MC (with error bars), HNC, Cohen, and  Ortner approximations.}

 \begin{tabular}{|l|l|l|l|l|} 
\hline
 $\Gamma$ &  $MC$  &    $HNC$   &    $Cohen$  &  $Ortner$     \\ \hline
$0.1$&$0.25117(34)$&$0.25688548$&$0.25677226$&$0.25699174$  \\ \hline
$0.2$&$0.34111(17)$&$0.34238929$&$0.34127338$&$0.34436859$  \\ \hline
$0.3$&$0.397693(64)$&$0.39837711$ &$0.39608173 $ &$0.40761777 $       \\ \hline
$0.4$&$0.439323(53)$&$0.43968253$ & $0.44115547 $  &$0.46432208 $                \\ \hline
$0.5$&$0.472172(42) $&$0.47208481$ & $0.49302326 $ &$0.521520956$   \\ \hline
$0.6$&$0.498715(21) $&$0.49850618$ & $0.57144385 $ &$0.58565711 $ \\ \hline
$0.7$&$0.521064(20) $&$0.52064202$ & $0.70120487 $ &$0.67244493 $  \\ \hline
$0.8$&$ 0.540173(15)$&$0.53956586$ & $0.91276540 $ &$0.81996338 $  \\ \hline
$0.9$&$0.556823(30) $&$0.55600050$ & $ 1.2425017$&$1.1053739$  \\ \hline
$1.0$&$0.571403(24) $&$0.57045534$ & $1.7327877 $ &$1.6651877$ \\ \hline
\end{tabular}
\end{table*}

\section{MC data analysis and fits}
\label{MCdat}
\subsection{Data analysis}
\label{data}
We adopted the same procedure as the one  described in reference I. 
The MC simulations were performed using the standard Metropolis algorithm to build Markov chains in the 
canonical ensemble.
In the small $\Gamma$ regime, $0 \leq \Gamma  \leq 1$, where finite size effects are tremendously important,
we considered  much larger systems than before. In order to get the thermodynamic limit (TL) of
the  excess internal energy for each value of $\Gamma$,  we performed simulations for samples of 
$N=100, 200, 400, 800, 1600, 3200, 6400, 12800, 25600, \textrm{ and } 51200$ particles.
The cumulated reduced excess  energy (CREE)  $\beta U(\Gamma, N)/\Gamma$ 
at coupling $\Gamma$ 
and number of particles $N$, was computed as the cumulated mean over $M$ successive configurations $"i"$ of the 
Markov chains as
\begin{equation}
\frac{\beta u_{N,\Gamma}(M)}{\Gamma}=
\frac{1}{M}\sum_{i=1}^{M}\frac{\beta V(i)}{N\Gamma} 
\; \; ( 1 \leq  M \leq \textrm{n}_{\textrm{nconf}}                  ) \; , \label{phg}
\end{equation}
We generated MC chains of  $\textrm{n}_{\textrm{nconf}} = 4. 10^9$ configurations after thermal equilibration,
for all systems  up to $N=25600$ particles. The reason was to 
to reach a stable plateau for the CREE and to reduce statistical errors. These two points will be illustrated further. For $N=25600$
such long chains result in the mixing of 5 independent chains, each one corresponding to half a month of CPU time. 
Thus the $N=25600$ value of the excess energy represent a 2 months and a half calculation. For  $N=12800$
the total duration was 2 monthes, with two independent chains. For comparison a  $N=800$ simulation is performed in 2 
days in a unique chain. One day is enough for a  $N=400$ simulation. These calculations have been performed 
simultaneously on the CEA Opteron clusters, local PC and the CRI cluster of Orsay, using one processor by job. 

In order to compute MC {\em statistical} errors on $\beta
u_{N}(\Gamma)/\Gamma$ each total run was divided into $n_{B}$ blocks and the error bar was obtained by a standard
block analysis \cite{Frenkel}. Each block involved a large number
$n_{B}^{conf}$ of successive MC configurations and was supposed to be 
statistically independent of the others.  For each calculation we checked that the variance was independent
of the size of the blocks for sufficiently large values of $n_{B}^{conf}$.
Results are so stable that we shall no more discuss this point in this paper.
The need of large simulations with $N=51200$ particles appeared with the difficulty to reach the thermodynamic limit and
to obtain the wanted precision for the $\Gamma$ values that we considered.
But, due to huge demand in CPU time of these simulations (one month for $10000$ configurations) only short chains were
considered, however long enough to reach the stable plateau of the CREE and to improve the TL research (see below).
Our data for $\beta u_{N}(\Gamma)/\Gamma$ are reported in table\ \ref{energie} where  the number 
in bracket  correspond to {\em one standard deviation $\sigma$}  and represent the accuracy of the last digits and only
the results for $N\geq 1600$ are given.
\begin{table*}
\caption{\label{energie} Minus the MC energy $\beta u_{N}(\Gamma)/\Gamma$ of
the OCP as a function of $\Gamma$ and  the 
number of particles $N$. The number in bracket which corresponds to {\em one
standard deviation $\sigma$} is the accuracy of the last digits.}
\begin{tabular}{|l|l|l|l|l|l|l|} 
\hline
$\Gamma$ &      $N=1600$    &     $N=3200$     & $N=6400$  &$N=12800$ & $N=25600$ &   $N=51200$          \\ \hline
$0.1$&$0.20942(7)$&$0.21343(8)$&$0.21984(8)$&$0.22728(7)$&$0.23436(7)$&$0.24018(18)$           \\ \hline
$0.2$&$0.32066(7)$&$0.32477(6)$&$0.32865(4)$&$0.33223(4)$&$0.33502(4)$&$0.337244(74)$            \\ \hline
$0.3$&$0.385447(36)$&$0.388389(23)$&$0.391016(23)$&$0.393158(27)$&$0.394704(25)$&$0.395825(99)$             \\ \hline
$0.4$&$0.430965(23)$&$0.433233(23)$&$0.435009(19)$&$0.436452(19)$&$0.437507(18)$&$0.438230(63)$             \\ \hline
$0.5$&$0.465821(16)$&$0.467568(17)$&$0.468939(17)$&$0.470015(17)$&$0.470754(15)$&$0.471263(52)$            \\ \hline
$0.6$&$0.493812(13)$&$0.495220(13)$&$0.496352(16)$&$0.497158(13)$&$0.497714(13)$&$0.498072(37)$              \\ \hline
$0.7$&$0.517109(13)$&$0.518254(12)$&$0.519178(12)$&$0.519806(11)$&$0.520259(13)$&$0.520600(32)$            \\ \hline
$0.8$&$0.536909(8)$&$0.537854(7)$&$0.538606(9)$&$0.539140(11)$&$0.539513(11)$&$0.539745(25)$             \\ \hline
$0.9$&$0.554034(10)$&$0.554810(12)$&$0.555458(8)$&$0.555886(11)$&$0.556232(10)$&$0.556458(40)$          \\ \hline
$1.0$&$0.569012(15)$&$0.569714(9)$&$0.570281(8)$&$0.570669(10)$&$0.570930(9)$&$0.571119(24)$\\ \hline
\end{tabular}
\end{table*}

\subsection{Connection with former simulations for  $1 \leq \Gamma  \leq 10$ }
\label{prec}
\begin{table}
\caption{Comparison with previous results of  the MC  energy
$\beta u_{N}(\Gamma)/\Gamma$ of the OCP in space ${\cal S}^{3}$ in function of 
the number of particles $N$ for $\Gamma=5$ and $\Gamma=10$. The first row, case "a",
corresponds to present study and second row, case "b", to table 1 of \cite{Caillol2}.
The only difference between the two calculations is the number of configurations,
typically \mbox{$n_{conf}=800 \, 10^{6}$} MC configurations after
equilibration in case "a", and \mbox{$n_{conf}= 5.10^{9}$} in case "b". 
The number in bracket which corresponds to {\em one standard deviation $\sigma$}
is the accuracy of the last digits. With two standard deviations the agreement 
is fulfilled.}
\label{energies}
\begin{tabular}{|r|r|r|r|r|r|r|} 
\hline
$\Gamma$ & $N=400$ &
  $N=800$ & $N=1600$ & $N=3200$ & $N=6400$ & $case$   \\ \hline
$5$    & $ $ &
         $ .7510930(37)$& $ .7511501(31)$& $ .7512037(35)$ & $.7512332(21)$& $a$\\ \hline
$5$    & $ .7510201(89)$ &
         $ .7511042(126)$& $ .7511513(135)$& $ .7511775(85)$ & $ $ & $b$ \\ \hline
$10$   & $ $ & 
         $.7998396(26)$ & $.7998148(30)$ & $.7998098(28)$ & $.7998043(15)$ & $a$\  \\ \hline                            
$10$   & $ .7998865(53)$ & 
         $ .7998414(43)$ & $ .7998149(51)$ & $ .7998131(55)$ & $ $ & $b$ \\ \hline                            
\end{tabular}
\end{table}
Before we present our new  results for $0 \leq \Gamma  \leq 1$, we shall study  the connection
with the results obtained in paper I, calculated
with the same MC code, but another range of $\Gamma$ values, i.e. $\Gamma  \geq 1$.
The only difference between the two calculations, calculated in double precision on 64 bytes work stations,
is thus the maximum number of configurations,
typically \mbox{$\textrm{n}_{\textrm{conf}}=800 \, 10^{6}$} MC configurations -after
equilibration in previous case (case "a"), and \mbox{$\textrm{n}_{\textrm{conf}}= 5.10^{9}$} in this paper (case "b"). 
We have performed comparisons for $\Gamma=1, 2, 3, 4, 5$ and $\Gamma=10$. 
The choices retained in I  were, at that time, the maximum reasonable conditions for the simulations.

\begin{figure}
\includegraphics[angle=0,scale=0.4]{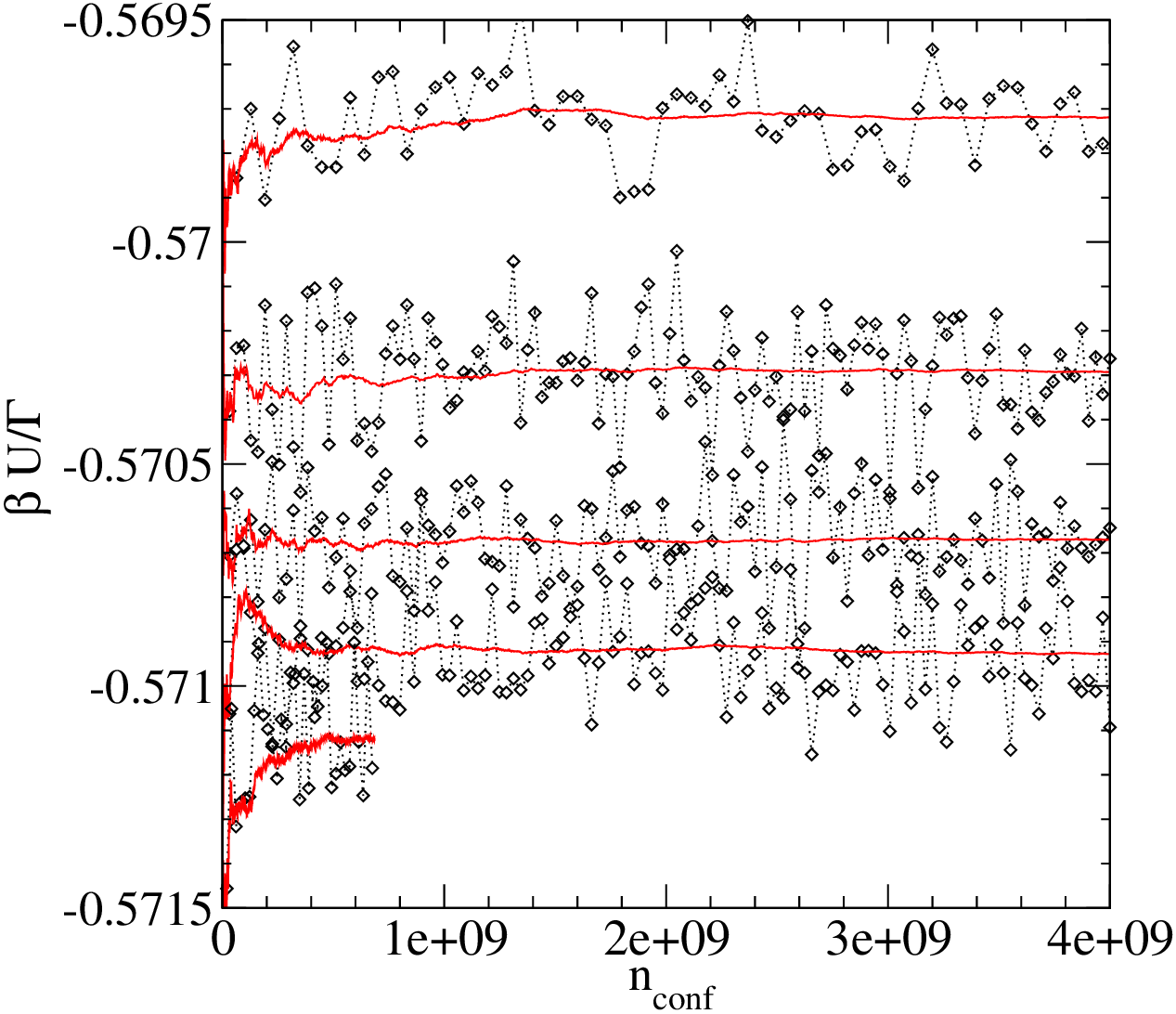}
\caption{\label{des_G=1} 
Solid lines: cumulated reduced excess energy $\beta u_{N}(\Gamma)/\Gamma$ versus the number of
configurations for $\Gamma=1$. From top to bottom $N=3200, 6400, 12800, 25600, 51200$.
Symbols: block averages.}
\end{figure}

Finite size effects decrease with increasing value of  $\Gamma$. Details of CREE's for $\Gamma=5$ and $\Gamma=10$
are reported in Table \ref{energies} ($N$ dependence of  MC  energy 
$\beta u_{N}(\Gamma)/\Gamma$ in both calculations, present and I).
For $\Gamma=10$ the results are in good agreement. 
But for $\Gamma=5$ slight discrepancies observed at $N=1600$ and $N=3200$ between the 
two calculations are a bit worrying. Indeed in these cases the error bars intervals do not overlap.
The main reason is that, for  the lowest $\Gamma$ results of ref.\ I.
the plateau of the CREE was in fact not reached.
\begin{figure}
\includegraphics[angle=0,scale=0.4]{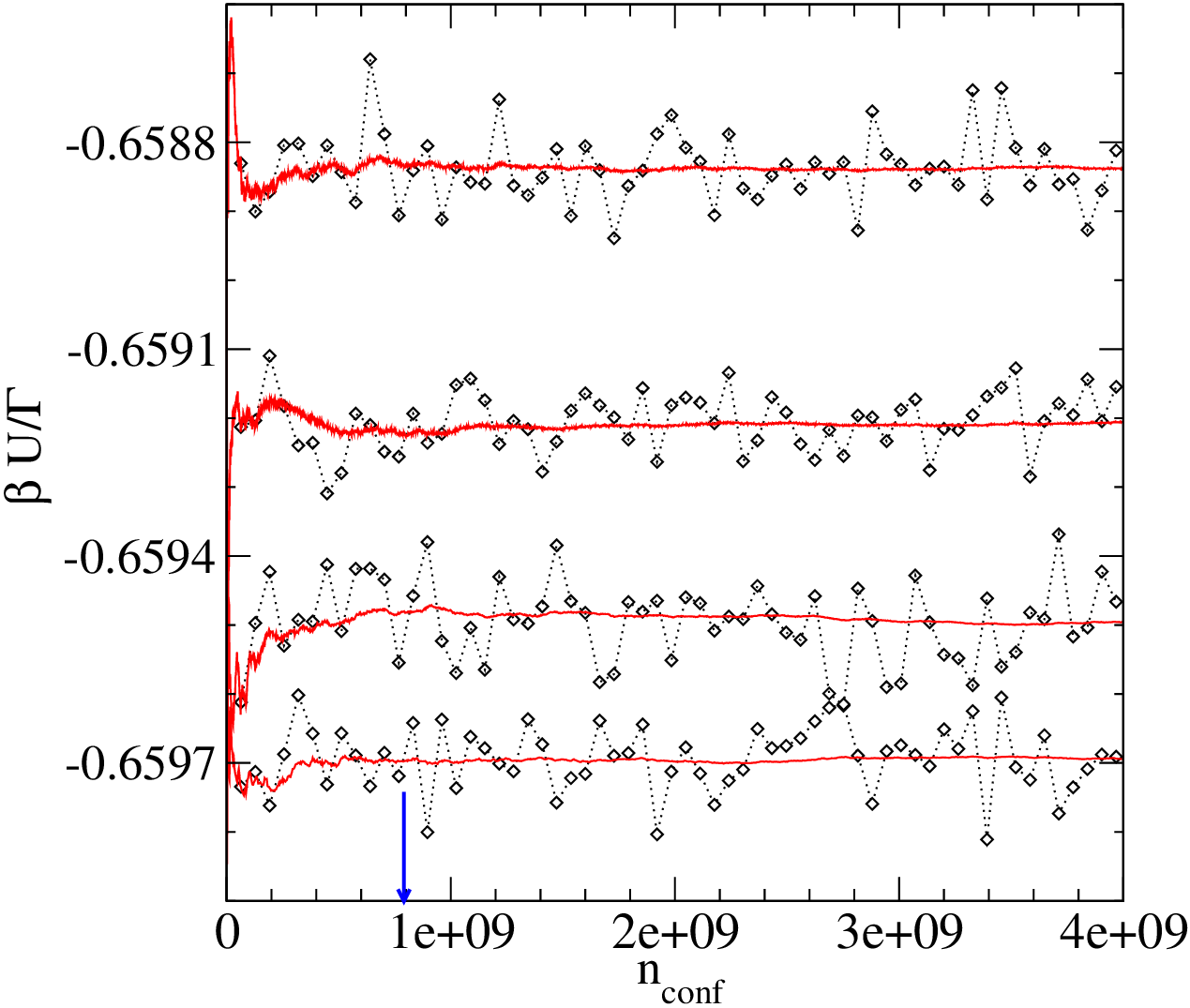}
\caption{\label{des_G=2} 
Solid lines: cumulated reduced excess energy $\beta u_{N}(\Gamma)/\Gamma$ versus the number of
configurations for $\Gamma=2$. From bottom to top $N=800, 1600, 3200, 6400$.
Symbols: block averages. The blue arrow points to the maximum number
of configurations $\text{n}_{\text{conf}}=8. \, 10^{8}$ considered in ref.~I.}
\end{figure}
This feature is illustrated by figure\ \ref{des_G=2} where
the CREE's for $\Gamma=2$ are displayed.
The figure illustrates the lack of configurations in the simulations of ref.~I for the CREE
$\beta U/\Gamma$ versus the number of configurations, displayed for different number of particles.
From top to bottom $N=800, 1600, 3200, 6400$. The blue arrow points on the maximum number of configurations
considered in I. When compared to our new calculations, clearly
the Markov chain  was  not long enough  to reach a plateau and such a drift of the CREE was 
probably underestimated in our previous calculations. The large variation with $N$ of the CREE with $N$
( solid red line )
gives an idea of the amplitude of finite size effects.
The simulation for the case $N=6400$, not included in paper I, has been added to improve the TL 
extrapolation. 
Only the $4.\, 10^{9}$ first configurations are plotted for visibility, but clearly each CREE value reaches its equilibrium value 
for a fixed $N$ value.
Figure \ref{des_G=1}  illustrates how previous conclusions for  the  case $\Gamma=2$ are emphasized
in the case $\Gamma=1$. 
It follows from the above remarks that a re-analysis of the TL of the energy of the OCP is necessary for 
$\Gamma =1,2, \ldots 10$. We recall the conclusions of I according to which the scaling law\ \ref{scaling} 
is valid only for low $\Gamma$ and that for $\Gamma \gtrsim 3-4$ the thermodynamic limit 
is reached more quickly with a scaling law 
\begin{equation}
\label{scaling_bis}
u_{N}\left( \Gamma \right) - u_{\infty}\left( \Gamma \right) \sim N^{-1} \; \; \mathrm{ for } \; \Gamma \gtrsim 3-4   \; \mathrm{ and }\;
\; N \to \infty \; .
\end{equation}
Moreover the scaling limits\ \ref{scaling} and\ \ref{scaling_bis} are satisfied, depending on the value of $\Gamma$, for very
large, and sometimes prohibitive large, numbers of particles $N$. The ideal case would be a linear fit passing through 
all the points within the error bars. This situation was indeed observed by including simulations at $N=51200$ particles 
and for not too low values of $\Gamma$. In other situations we had to content ourselves with quadratic fits including 
the next leading order term (i.e. either $\mathcal{O}(1/N^{2})$ or $\mathcal{O}(1/N^{4/3})$ according to the value of 
$\Gamma$).
In Table  \ref{limhigh} are resumed the comparisons for the TL of the energy
between present and results of paper I for $\Gamma=1, 2, 3, 4, 5, 10$. The type of the extrapolation scheme is specified
in the column "fit", together with 
the interval of $N$ values considered for the fit. Precision are also reported.
For $\Gamma \le 5$  it is clear that results are slightly shifted between calculations. As expected for $\Gamma=10$ present
and previous results are similar; higher values of $\Gamma$ should not cause any trouble.

\begin{table}
\caption{\label{limhigh} Thermodynamic limit of the energy of the OCP 
versus $\Gamma$ for $\Gamma  \geq 1$ of , case "a", compared to previous calculations,
case "b". The difference between two calculations is the maximum number of partparticles
(no more than $3200$ in case a) and the total number of configurations.
The type of extrapolation scheme is specified in the column "fit".
 For instance $quad(3200-51200)$ means
that a quadratic regression involving the data from $N=3200,6400,12800,51200$ has been used.
The variable entering the fit is specified in the column Variable.
p is the precision of the fit.
The number in bracket which corresponds to {\em one
standard deviation $\sigma$} is the accuracy of the last digits.}

\begin{tabular}{|l|l|l|l|l|l|l|l|} 
\hline
$\Gamma$ & $\beta u_{\infty}/\Gamma$& $Fit$&$ p*10^{5}$&$\beta u_{\infty}/\Gamma$& $Fit$& $p*10^{5}$&$Variable$\\ \hline
$   $     &        $ a $     &     $ a $  &  $ a $  &  $b$  &    $b$  &    $b$  &     \\ \hline        
$1.$&$-0.571387(24)$&lin(12800-51200)& $4.2$ &$-0.571098(39)$&cub(100-3200) &$6.9$&$N^{-2/3}$ \\ \hline
$1.$&$-0.571403(22) $&quad(3200-51200)& $3.8$&$             $&              &   &$N^{-2/3}$\\ \hline
$2.$&$-0.6598934(68)$&quad(800-6400)&$7.0$ &$-0.659983(23)$&quad(200-3200)&$3.5$&$N^{-2/3}$\\ \hline
$3.$&$-0.7042987(54)$&quad(3200-6400&$0.8$ &$-0.704348(19)$&quad(200-3200)&$2.7$&$N^{-2/3}$\\ \hline
$4.$&$-0.7319760(46)$&lin(800-6400)&$0.6$&$-0.731916(12)$&quad(200-3200)&$1.7$&$N^{-1}$\\ \hline
$5.$&$-0.7512608(22)$&lin(800-6400)&$0.3$&$-0.7512126(98)$&quad(200-3200)&$1.3$&$N^{-1}$\\ \hline
$10.$&$-0.7997991(16)$&lin(800-6400)&$0.2$ &$-0.7997974(45)$&lin(400-3200) &$0.56$&$N^{-1}$\\ \hline
\end{tabular}
\end{table}


\subsection{Thermodynamic limit extrapolation scheme}
\label{schem}
The aim of our simulations was to compute the TL of the energy $\beta u_{N=\infty}(\Gamma)$ with a high degree of
accuracy by taking into
account finite size effects which are of overwhelming importance  for $\Gamma \leq 1$. The need of simulations
up to $N=51200$  and involving  no less than $N = 800$, or even $N = 1600$ particles for the smallest
values of $\Gamma$, appeared crucial to reach the scaling law\ \ref{scaling}. It appears that, for this range
of $N$, MC data can be fitted with the quadratic fits
\begin{equation}
\label{fit_qua}
\beta u_{N}(\Gamma) =\beta u_{N=\infty}(\Gamma) +a_1 \frac{1}{N^{2/3}} + a_2 \left[ \frac{1}{N^{2/3}}\right] ^{2} \, .
\end{equation}
For most values of $\Gamma$ it proved possible to explicitely check the asymptotic form linear in $N^{-2/3}$ (i.e. $a_2=0$ in
equation\ \ref{fit_qua}) by keeping only the 3 largest systems, i.e. $N=12800, 25600$ and $N=51200$.
Recall that in paper I the largest considered systems were made of $N=3200$ particles. An 
exhaustive discussion follows in next section.

\subsection{Results for $0 \le \Gamma \le1$}
\label{mcvew}

\begin{figure}
\includegraphics[angle=0,scale=0.4]{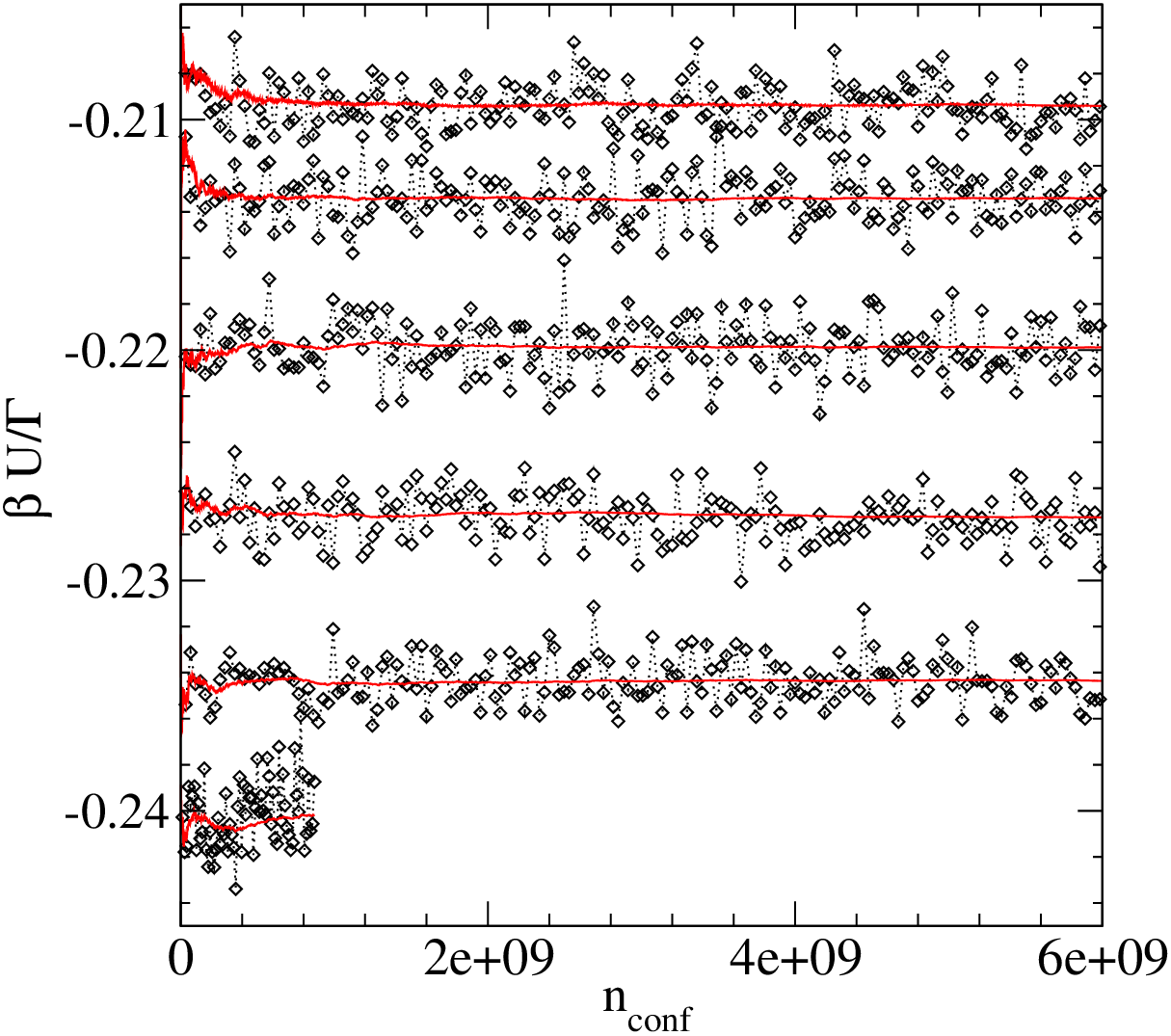}
\caption{\label{des_G=01} 
Solid lines: cumulated reduced excess energy $\beta u_{N}(\Gamma)/\Gamma$ versus the number of
configurations for $\Gamma=0.1$. From top to bottom $N=1600, 3200, 6400, 12800, 25600, 51200$.
Symbols: block averages.}
\end{figure}

\begin{figure}
\includegraphics[angle=0,scale=0.4]{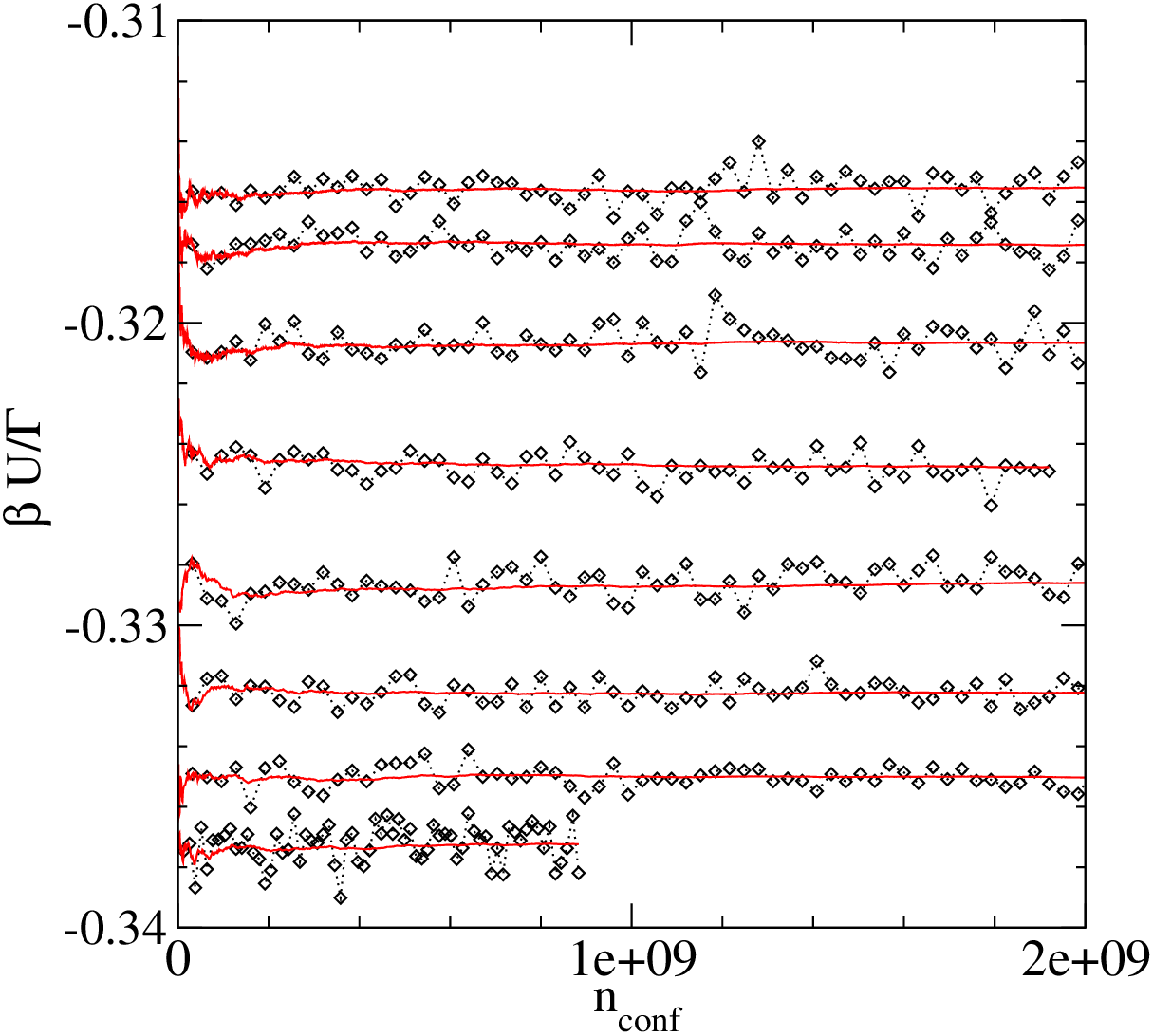}
\caption{\label{des_G=02} 
Solid lines: cumulated reduced excess energy $\beta u_{N}(\Gamma)/\Gamma$ versus the number of
configurations for $\Gamma=0.2$. From top to bottom $N=400, 800, 1600, 3200, 6400, 12800, 25600, 51200$.
Symbols: block averages.}
\end{figure}

\begin{figure}
\includegraphics[angle=0,scale=0.4]{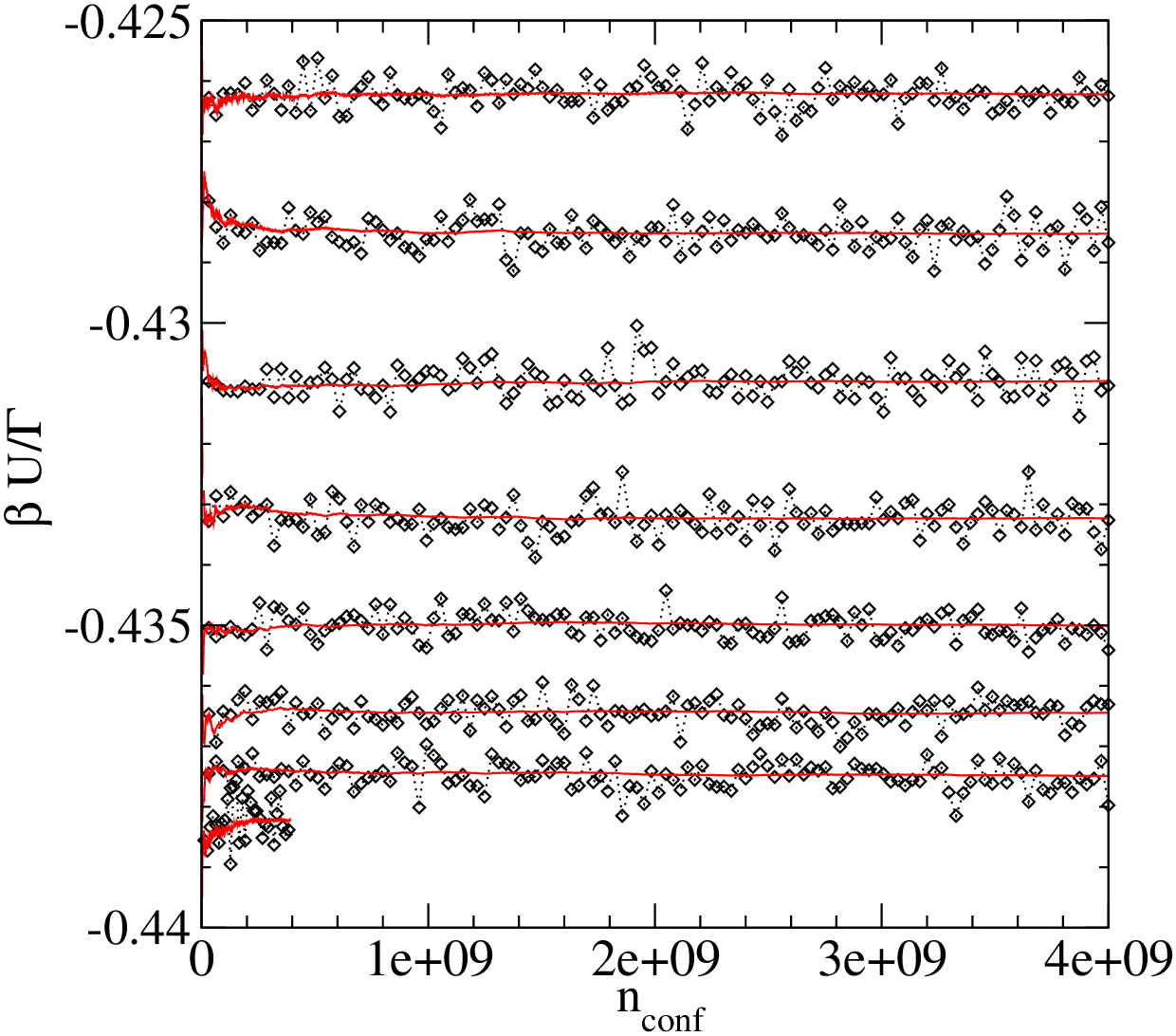}
\caption{\label{des_G=04} 
Solid lines: cumulated reduced excess energy $\beta u_{N}(\Gamma)/\Gamma$ versus the number of
configurations for $\Gamma=0.4$. From top to bottom
 $N=400, 800, 1600, 3200, 6400, 12800, 25600, 51200$.
Symbols: block averages. }
\end{figure}

\begin{figure}
\includegraphics[angle=0,scale=0.4]{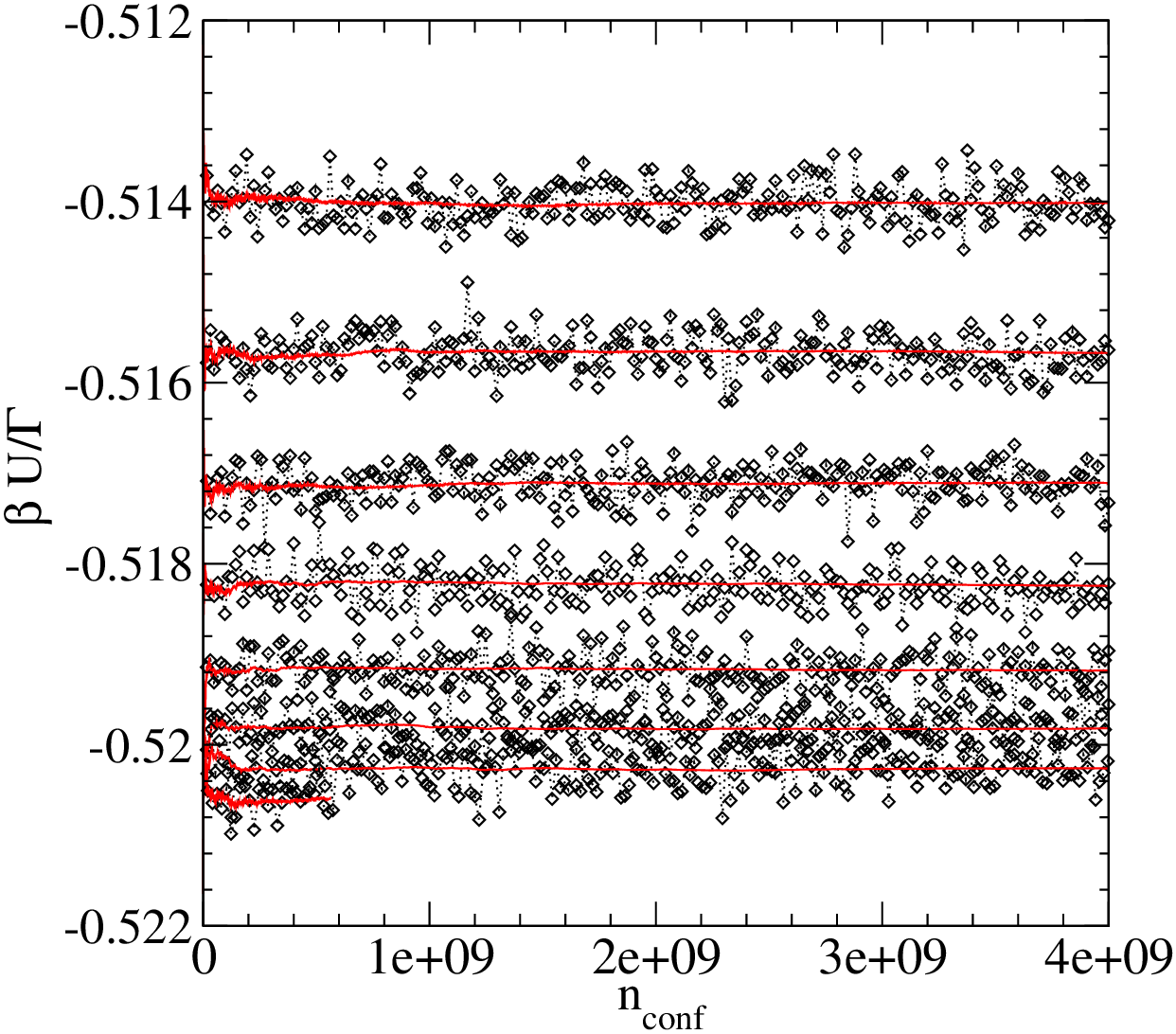}
\caption{\label{des_G=07} 
Solid lines: cumulated reduced excess energy $\beta u_{N}(\Gamma)/\Gamma$ versus the number of
configurations for $\Gamma=0.7$. From top to bottom $N=400, 800, 1600, 3200, 6400, 12800, 25600, 51200$.
Symbols: block averages.}
\end{figure}

\begin{table}
\caption{\label{lim} Thermodynamic limit of the energy of the OCP 
versus $\Gamma$. The number in bracket which corresponds to {\em one
standard deviation $\sigma$} is the accuracy of the last digits. 
The type of extrapolation scheme is specified in the column "fit".
The variable entering the fit is $N^{-2/3}$.}
\begin{tabular}{|l|l|l|} 
\hline
$\Gamma$ & $\beta u_{\infty}/\Gamma$& Fit \\ \hline
$0.1$&$-0.25117(34)$&quad(6400-51200)  \\ \hline
$0.2$&$-0.34111(17)$&quad(6400-51200)  \\ \hline
$0.3$&$-0.397693(64)$&quad(3200-51200)  \\ \hline
$0.4$&$-0.439323(53)$&lin(12800-51200)  \\ \hline
$0.4$&$-0.439528(50)$&quad(3200-51200)  \\ \hline
$0.5$&$-0.472028(45)$&lin(12800-51200)  \\ \hline
$0.5$&$-0.472172(42)$&quad(3200-51200)  \\ \hline
$0.6$&$-0.498663(38)$&lin(12800-51200)  \\ \hline
$0.6$&$-0.498715(21)$&quad(1600-51200)  \\ \hline
$0.7$&$-0.521063(32)$&lin(12800-51200)  \\ \hline
$0.7$&$-0.521064(20)$&quad(1600-51200)  \\ \hline
$0.8$&$-0.540146(28) $&lin(12800-51200)  \\ \hline
$0.8$&$-0.540173(15) $&quad(1600-51200)  \\ \hline
$0.9$&$-0.556823(30)$&lin(12800-51200)  \\ \hline
$0.9$&$-0.556801(25) $&quad(3200-51200)  \\ \hline
$1.0$&$-0.571387(24)$&lin(12800-51200)  \\ \hline
$1.0$&$-0.571403(22) $&quad(3200-51200)  \\ \hline
\end{tabular}
\end{table}
\noindent
We present and discuss in details the ten values  $\Gamma=0.1, 0.2, \ldots ,1$ considered in our
numerical experiments.
Figures \ref{des_G=01}, \ref{des_G=02}, \ref{des_G=04} and \ref{des_G=07} illustrate the CREE
$\beta u_{N}(\Gamma)/\Gamma$ versus the number of configurations for  several caracteristic  values of
$\Gamma$( $\Gamma=0.1$, 0.2, 0.4, and 0.7 respectively)
 typical of the different plasma regimes in the interval $(0,1)$.
Our previous comments on figures \ref{des_G=1} and \ref{des_G=2} (for $\Gamma=1, 2$ respectively)
are still valid in these cases. We stress once again
the need of large systems together with the need of enough  configurations to reach a stable plateau after 
thermal equilibration. 

All generated configurations, $n_{conf}=6.10^{9}$, are displayed in 
figure \ref{des_G=01} ($\Gamma=0.1$) while a zoom of only the first $2.\, 10^{9}$ configurations is displayed in
figure \ref{des_G=02} ($\Gamma=0.2$), which exemplifies the plateau reached by the CREE for $N=51200$.
On the last two figures  \ref{des_G=04} and \ref{des_G=07}, respectively 
for $\Gamma=0.4$ and $\Gamma=0.7$ and $n_{conf}=4.\,10^{9}$,
we see  the good convergence with $N$ as the interval width between CREE values decreases from top to bottom. 
By contrast the low $\Gamma$ runs do not exhibit this regular decrease. Of course beyond visual impressions 
only the possibility and precision of the fitting process 
of the MC CREE results will give a firm answer on the quality of the TL calculation for each $\Gamma$ value.

\begin{figure}
\includegraphics[angle=0,scale=0.4]{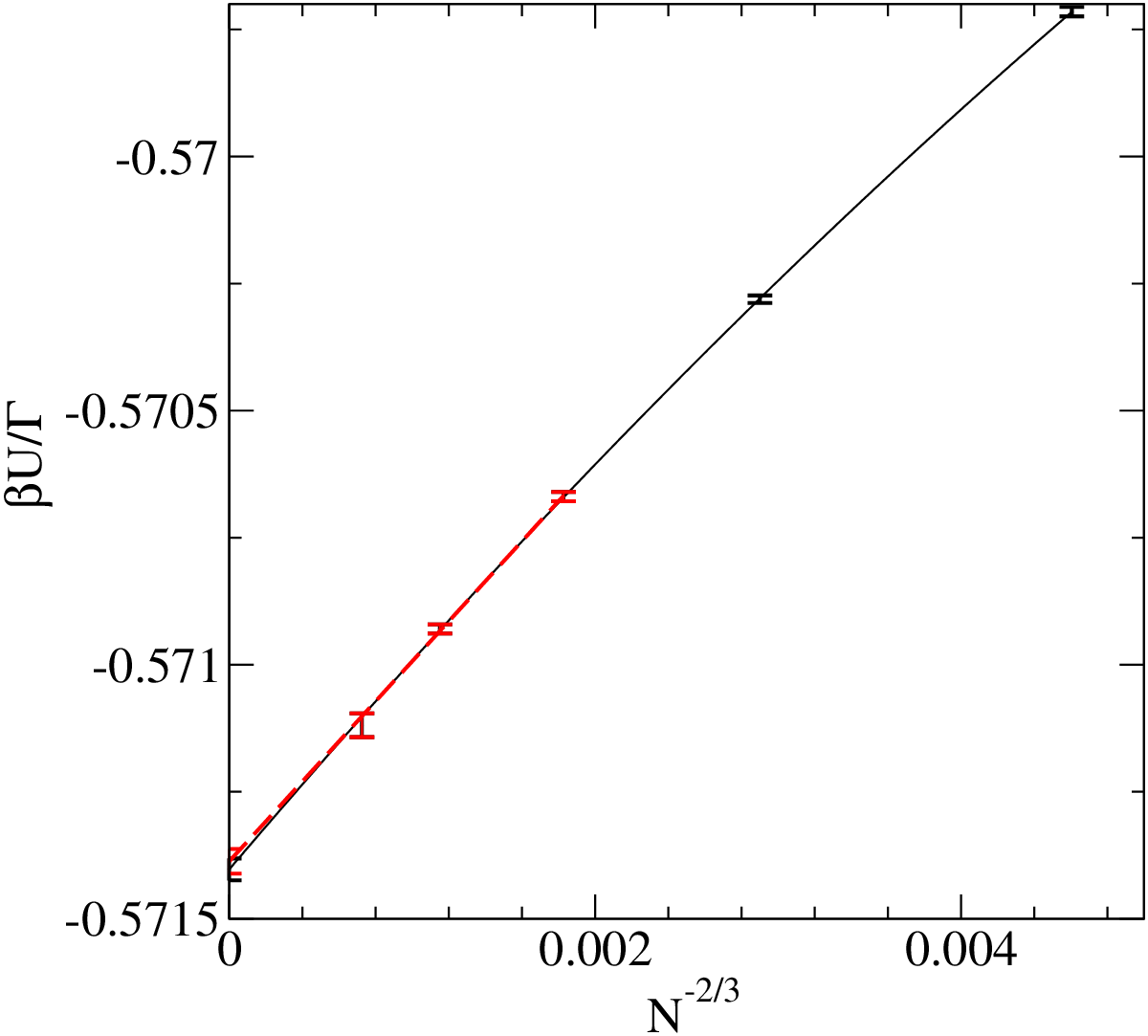}
\caption{\label{fit_G=1} 
Solid lines: cumulated reduced excess energy $\beta u_{N}(\Gamma)/\Gamma$ versus $1/N^{2/3}$
for $\Gamma=1$. From left to right $N=\infty,51200,25600,12800,6400,3200$. The error bars
correspond to one standard deviation $\sigma$.  Solid black line : quadratic polynomial regression  of  MC data.
Dashed red line : linear regression  of the 3 larger systems MC data.}
\end{figure}

\begin{figure}
\includegraphics[angle=0,scale=0.4]{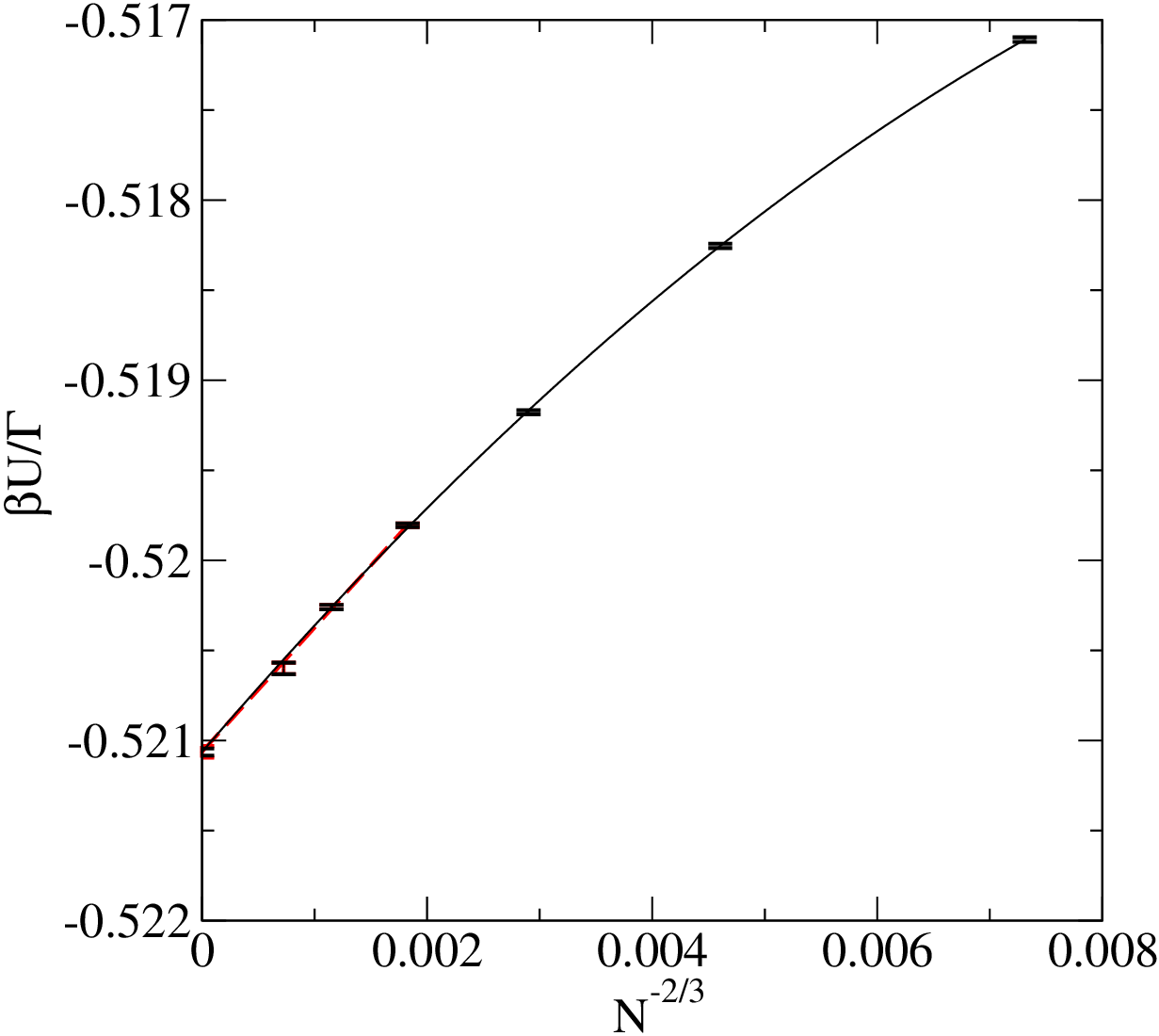}
\caption{\label{fit_G=07} 
Same legend than figure\ \ref{fit_G=1} but for $\Gamma=0.7$. From left to right 
$N=\infty,51200,25600,12800,6400,3200,1600$.}
\end{figure}

\begin{figure}
\includegraphics[angle=0,scale=0.4]{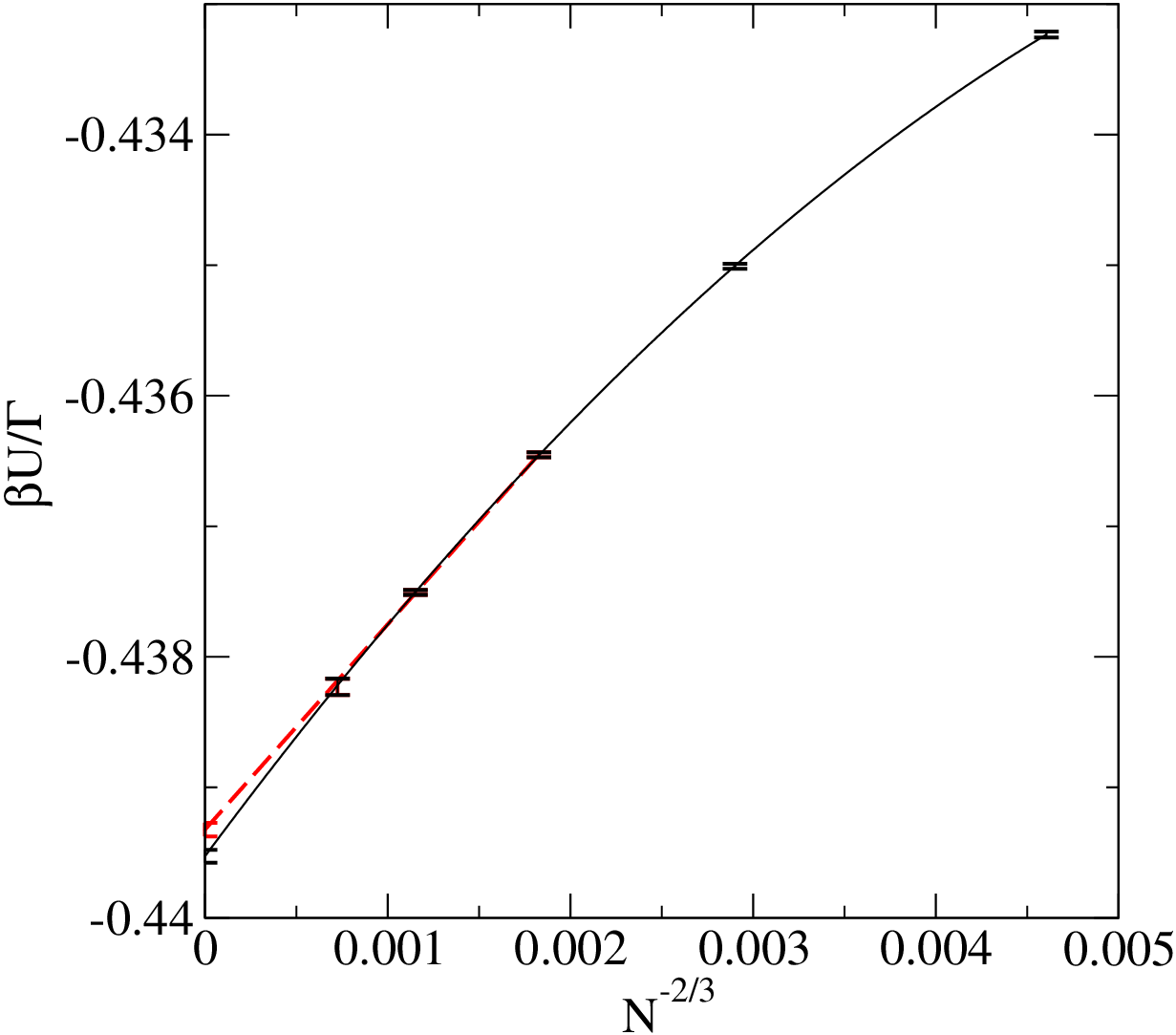}
\caption{\label{fit_G=04} 
Same legend than  figure\ \ref{fit_G=1} but for $\Gamma=0.4$. From left to right 
$N=\infty,51200,25600,12800,6400,3200$.}
\end{figure}

\begin{figure}
\includegraphics[angle=0,scale=0.4]{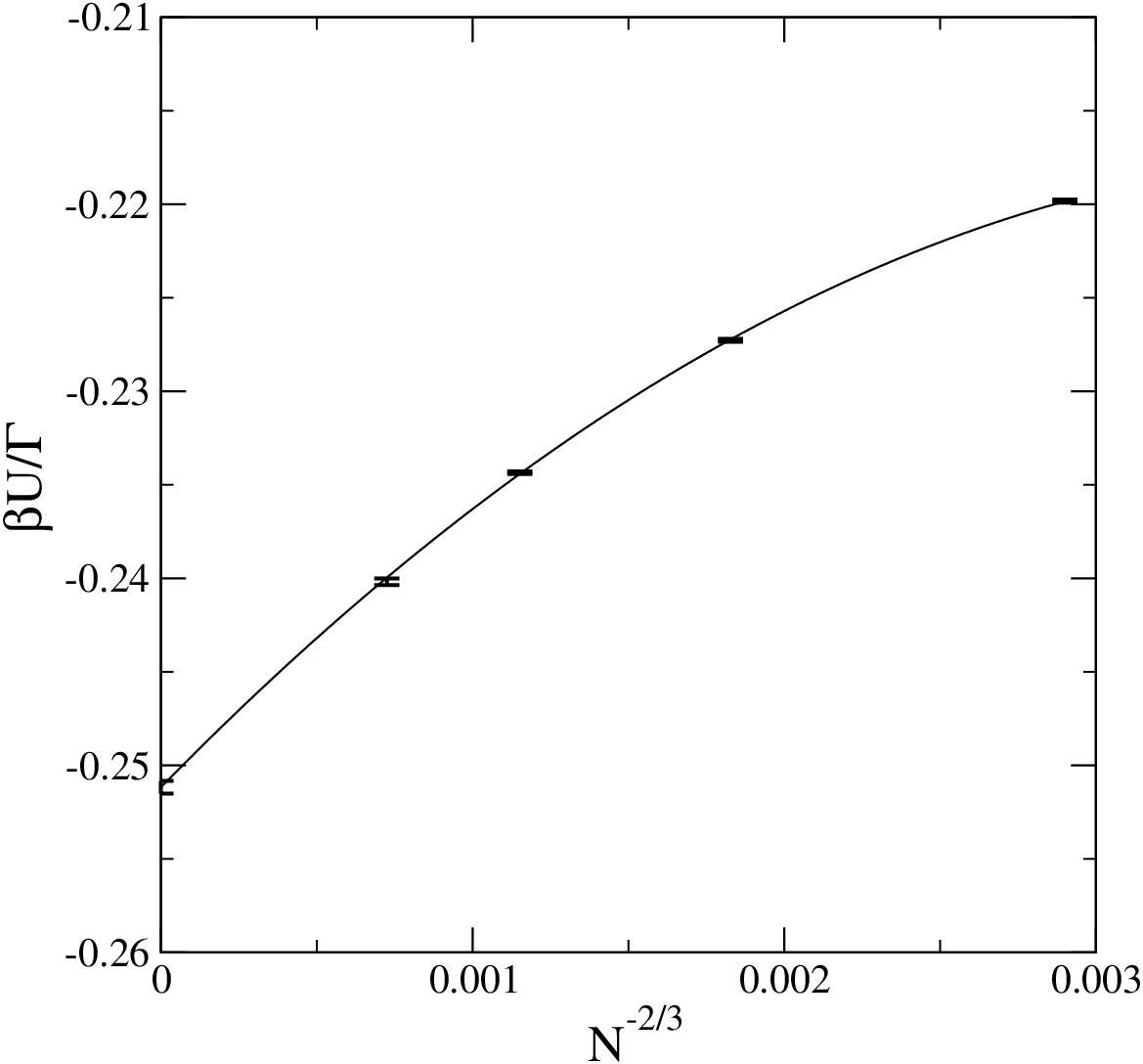}
\caption{\label{fit_G=01} 
Solid lines: cumulated reduced excess energy $\beta u_{N}(\Gamma)/\Gamma$ versus $1/N^{2/3}$
for $\Gamma=0.1$. From left to right $N=\infty,51200,25600,12800,6400,3200$. The error bars
correspond to one standard deviation $\sigma$.
Solid black line :  quadratic polynomial regression  of  MC data.
}
\end{figure}

Table \ref{energie} resumes present work MC calculations of  the MC energy $\beta u_{N}(\Gamma)/\Gamma$ of
the OCP as a function of $\Gamma$ for $N=1600$ to  $N=51200$.
The number in bracket which corresponds to {\em one standard deviation $\sigma$} is the accuracy of the last digits. 
Results corresponding to $N\le1600$, not included in the fits, are not reported.
The thermodynamic limit values of the energy 
versus $\Gamma$ are reported in Table \ref{lim}. The type of extrapolation schemes retained in the fits, 
i.e. linear or quadratic (cf equation\ \ref{fit_qua}), are specified. 
In figures \ref{fit_G=1}, \ref{fit_G=07}, \ref{fit_G=04} and \ref{fit_G=01} we display the quadratic fit of 
$\beta u_N(\Gamma)/ \Gamma$ (solid black line) and the linear fits for the 3 largest numbers of particles
considered, when available  (red dashed line) for  $\Gamma=1$,  0.7, 0.4,  and $\Gamma=0.1$ respectively.
The error bars on the value of the TL of the energy  $\beta u_{\infty}(\Gamma)/ \Gamma$ reported
in table\ \ref{lim} are the error bars of the linear (or quadratic) regression.

We discuss now the results from high to low $\Gamma$'s. 
Figure\ \ref{fit_G=1}  illustrates the high quality of the fits obtained in
the case $\Gamma=1.0$. Indeed the extrapolated TL of  $\beta u_{N}(\Gamma)/\Gamma$ coincide for both the linear 
and the quadratic fits (the latter involving more states with low number of particles and the former only the 3 largest systems) 
with a nice overlap of the error bars.
Results are similar down to $\Gamma=0.7$, as illustrated in Figure \ref{fit_G=07} for $\Gamma=0.7$.

In the range $0.5 \le\Gamma \le 0.7$
the precision of the fits is good but  the linear and the quadratic fit extrapolations do not give exactly the same 
TL values, however the error bars do overlap. 
Figure\ \ref{fit_G=04}, corresponding to  the case $\Gamma=0.4$, illustrates the smallest   $\Gamma$ at which
a linear fit is possible with the 3 higher values of $N$.  The linear and the quadratic fit extrapolations giving
the TL values would coincide within the error bars if the latter were defined to be \textit{two standard deviations}
rather than only  \textit{one} according to our choice.

For $\Gamma$ smaller or equal to 0.3  it was impossible to reach an asymptotic form of
$\beta u_{N}(\Gamma)/\Gamma$,  linear in the variable 
$N^{-2/3}$, and only a quadratic polynomial fit  was possible (cf table\ \ref{lim}). For that reason it is legitimate
to consider the error bars on the extrapolated value $\beta u_{\infty}(\Gamma)/\Gamma$ as overoptimistic 
in this range of   $\Gamma$, see figure\ \ref{fit_G=01} for an illustration in the case $\Gamma=0.1$. Simulations
involving larger numbers of particles would be necessary but are out of our reach.

For all the states with a $\Gamma \geq 0.4$ the TL limit  $u_{\infty}(\Gamma)$ can thus be obtained with a high precision
$p \sim 10^{-5}$, after a careful study of finite size effects on the MC energies $u_{N}(\Gamma)$. For smaller values of
$\Gamma $, for instance $\Gamma =0.1$,  samples of more than $N\simeq 200000$ particles should be used
to reach the leading order of the asymptotic regime\ \ref{scaling}. However such an effort would be useless since
HNC and Cohen approximations are then "exact" within the wanted precision on $u$. 
The $u_{\infty}(\Gamma)$ are perfectly well fitted in the range $0.4 \leq \Gamma \leq 1$ by the Cohen's
functional form, given
by equation\ \ref{a}, involving the five parameters $p_i$ ($i=0, \ldots, 4$) given in table\ \ref{coef4}.

\section{Conclusion}
\label{Conclusion}
\noindent
In this conclusion we compare at first the Cohen-Ortner low $\Gamma$ expansions, HNC and MC data.
Figure \ref{fig_COMPARAISON_U}  shows without ambiguity the good agreement between HNC and MC results in the
range $ 0 \leq \Gamma \leq 1$ and the large departure of both results with analytical expansion ones, DH 
(for $ \Gamma \ge 0.05$), Th1 (for $ \Gamma \ge 0.3$) and Th2 (for $\Gamma \ge 0.2$). 
Note however that the  scale  of the figure is not large enough to discriminate between HNC and MC results, 
notably because the errors bars on MC results are smaller than the
size of the symbols. A more enlightening illustration is that of figure\ \ref{fig_RAP} which gives the ratio of the 
MC and HNC energies. The disagreement for $\Gamma \le 0.3$ results from a bad evaluation of the TL of $u_N$
due to huge finite size effects spoiling the MC data, while the disagreement for $\Gamma \ge 0.6$ 
simply reflects the failure of
the HNC approximation at high  $\Gamma$. A nearly perfect agreement between MC and HNC results, compatible with
\textit{one standard deviation} is observed only at $\Gamma = 0.5$; with \textit{two standard deviations}
the HNC results are within the error bars of the MC data in the interval  $ 0.4 \leq \Gamma \leq 0.6$.
By passing our new HNC calculations for some values of $\Gamma$ in the range $(1,10)$ are plotted
in figure \ref{fig_HORREUR_DE_HNC}  were MC data were also included for comparison.

It is the place to resume our analysis. We found that, for a wanted precision of $p=10^{-5}$ on the energy :
\begin{itemize}
\item $0 \leq \Gamma \leq 0.05$ is the range of validity of Debye-H\"uckel theory.
\item $0 \leq \Gamma \leq 0.3$ is the range of validity of Cohen low $\Gamma$ expansion \ \ref{eqnortner}.
\item Ortner's additional terms do not improve the results.
\item  $0 \leq \Gamma \leq 0.5$ is the range of validity of HNC. The data are perfectly represented by the eight parameters
           fit of tables\ \ref{coef4} and\ \ref{coef567}. 
\item We were able to extract the thermodynamic limit of the OCP energy from our MC simulations 
        with a precision not smaller than $p=10^{-5}$ in the range 
        $0.4 \leq \Gamma \leq 1$. Our data are well fitted by the five parameters fit of table\ \ref{coef4}.
\end{itemize}

\section*{}

\end{document}